\documentstyle[aps,prd,epsfig]{revtex}

\begin{document}

\newcommand{\be}{\begin{equation}}
\newcommand{\ee}{\end{equation}}
\newcommand{\bea}{\begin{eqnarray}}
\newcommand{\eea}{\end{eqnarray}}
\newcommand{\vp}{\varphi}
\def\vec#1{{\mathbf{#1}}}

\newcommand{\mcommand}[1]{\relax\ifmmode #1\else $#1$\fi}
\newcommand{\dRdOmega}{\mcommand{\frac{dR}{d\Omega_{\gamma,\phi}}}}
\newcommand{\vearth}[1]{\mcommand{v_{\oplus,#1}}}
\newcommand{\Eth}{\mcommand{E_{\rm th}}}
\newcommand{\vex}{\vearth{x}}
\newcommand{\vey}{\vearth{y}}
\newcommand{\vez}{\vearth{z}}
\newcommand{\unit}[1]{\mcommand{\;{\rm #1}}}
\newcommand{\kmps}{\unit{km/s}}
\newcommand{\kpc}{\unit{kpc}}
\newcommand{\GeV}{\unit{GeV}}
\newcommand{\keV}{\unit{keV}}
\newcommand{\Ge}{\mcommand{\vphantom{\rm Ge}^{73}{\rm Ge}}}

\draft

\title{Angular Signatures for Galactic Halo WIMP Scattering in
Direct Detectors: Prospects and Challenges}


\author{Craig J. Copi$^1$ and Lawrence M. Krauss$^{1,2}$
}
\address{$^1$Department of Physics, $^2$Department of Astronomy \\
Case Western Reserve University, 
10900 Euclid Ave., Cleveland OH 44106-7079}

\date{\today}


\maketitle

\rightline{{CWRU-P8-00}}

  
\begin{abstract}

Angular Sensitivity  can provide a
key additional tool which might allow unambiguous separation of a signal
due to Galactic halo WIMPs from other possible backgrounds in direct
detectors.  We provide a formalism which allows a calculation of the
expected angular distribution of events in terrestrial detectors with
angular sensitivity for any incident  distribution of Galactic
halo dark matter. This can be used as an input when studying the
sensitivity of specific detectors to halo WIMPs.  We utilize this
formalism to examine the expected signature for WIMP dark matter using a
variety of existing analytic halo models in order to explore how 
uncertainty in the Galactic halo distribution impact on the the
event rates that may be required to separate a possible WIMP signal from
other terrestrial backgrounds.  We find that as few as 30 events might
be required to disentangle the signal from backgrounds if the WIMP
distribution resembles an isothermal sphere distribution.
On the other hand, for certain halo
distributions, even  detectors with fine scale resolution may require in
excess of a 100-400 events to distinguish a WIMP signal from backgrounds
using angular sensitivity. We also note
that for finite thresholds the different energy dependence of
spin-dependent scattering cross sections may require a greater number of
events to discern a WIMP signal than for spin independent interactions.  
Finally, we briefly describe ongoing studies aimed at developing
strategies to better exploit angular signatures, and the use of N-body
simulations to better model the expected halo distribution in predicting
the expected signature for direct WIMP detectors. 

\end{abstract}




\section{Introduction}

The effort to directly detect Galactic halo weakly interacting
massive particles (WIMPs) which may form the dark matter dominating our
Galaxy is now entering a new phase.  Numerous experiments are now
operating with a sensitivity to scattering cross sections in the range
predicted for low energy SUSY neutralinos \cite{review,cdms,idm2000},
and indeed one such experiment has claimed a preliminary detection
\cite{dama}.

There is a problem however, with existing detection schemes, which the
controversy over the recent DAMA results underscores.   The signature for
elastic scattering of WIMPs on nuclear targets involves the
observation of what is essentially excess noise in the detector.  Thus, in
order to definitively establish a positive detection, one has to
convincingly demonstrate that this excess noise is indeed WIMP induced,
and not due to some unexpected terrestrial background.  The signature
focussed on by the DAMA collaboration, an annual modulation due
to the Earth's orbital motion through the WIMP halo, is 
problematic in this regard.  In the first place, the effect is small, at
the few percent level at best \cite{annual-variation}.  In addition,
numerous radioactive backgrounds are known to modulate over the course of
a year.

The Earth's motion around the Sun, combined with the Sun's motion around
the Galactic disk, can however provide another, much larger and more
unambiguous signature.   If detectors can be developed that are
sensitive to the direction of recoil of the nucleus impacted by a WIMP,
then the expected signal will have an angular dependence characteristic
of this velocity relative to the incident WIMP halo phase space
distribution.    

For a simple isothermal sphere distribution, the resulting angular
differential rates can be easily calculated \cite{spergel}, and it is
clear that a large forward-backward asymmetry can be produced.  
However, it is quite likely that the actual Galactic halo WIMP
distribution is not well represented by such a simple analytical 
approximation. Building detectors with angular 
sensitivity is a daunting challenge, although testing of at least one
such proposed detector, the DRIFT detector~\cite{angulardet,tpc} is
currently underway.  It is therefore appropriate to explore in advance to
what extent  uncertainties in our knowledge of a halo WIMP distribution
will have on the detector parameters required required to separate a
WIMP signal from other backgrounds. 

Recently we outlined a comprehensive formalism which
allows the generation of expected angular event rates in
terrestrial detectors with angular sensitivity for any incident WIMP
distribution, and made a provisional exploration of the overall event
rates required in detectors for simple halo models in order to separate
the signal from a flat background
\cite{copietal}.  Here, we follow up by providing a detailed
derivation and description of our results, and also by incorporating a much
broader spectrum of analytical halo models.  In addition, we also make a
provisional analysis of the possible effects of spin-dependence of
WIMP-nucleon cross sections on the resulting angular signatures. 
Finally, we briefly describe ongoing work aimed at incorporating more
realistic WIMP distributions arising from N-body simulations into our
analysis, and applying our results to various specific proposed
detectors, where energy-dependence of both signals and backgrounds may
play a key role.   Our present results, however, point out the challenges
of discerning an angular signal in advance of specific knowledge of the
halo WIMP distribution.  In addition, the formalism provided here can be
used to generate angular event rates which can in turn be used by
experimental groups as an input into their detector Monte Carlo analyses.

The outline of this work is as follows:  First, in section II, we present
a detailed derivation of the differential angular cross sections in a
terrestrial laboratory frame for any incident WIMP distribution.  In
section III we discuss a broad range of different analytical halo models
that have been proposed in the literature, and outline the relevant
parameter ranges of importance for influencing the resulting angular
signature for WIMP scattering.  In this section we also incorporate
effects of the detailed motion of the Earth in a Galactic-rest frame, and
also examine the effect of incorporating possible spin-dependence of
WIMP-nucleus scattering on the resulting angular signals, as well as
briefly outlining the generic detector and WIMP parameters we used in
generating our results.  In section IV we present the results for the
expected angular distributions in these model detectors.  In section V
we then proceed to describe the Monte Carlo analysis we performed to
determine the required event rates in order to separate signals from
backgrounds for the different halo and detector models.  In section VI
we summarize our conclusions, and describe ongoing efforts to further
refine our analyses and predictions.

\section{Differential Event Rates}

The WIMP rate as a function of recoil energy for various WIMP models has
been carefully studied over the past decade (see~\cite{review} for a
review).  The angular dependence of the event rate, assuming an incident
spherically symmetric isothermal halo WIMP distribution has also been
discussed.~\cite{spergel,smith}. In order to explore a more general class
of halo distributions, in particular those which might not be spherically
these formalisms are not adequate, however, and one must generalize
them \cite{copietal}.  We present the derivation of such a generalized
formalism here.

The event rate of WIMPs depends on their local density, $\rho_0$, and
their velocity distribution in the halo, $f (\vec{v})$.  Assume the
WIMPS have a uniform spatial density on the scale of the Earth's motion
through  the halo over the course of one year.  The
event rate is then simply given by 
\be R \sim \frac{n\sigma}{m_n} \left<v\right>, \ee
where $m_n$ is the mass of the nuclear target, $m_\chi$ is the mass of the
WIMP, $n=\rho_0/m_\chi$ is the number 
density of WIMPs, $\sigma$ is the cross section, and $\left<v\right>$ is the
average velocity of the WIMPs relative to the detector.

\begin{figure}[tbp]
  \leavevmode\center{\epsfig{figure=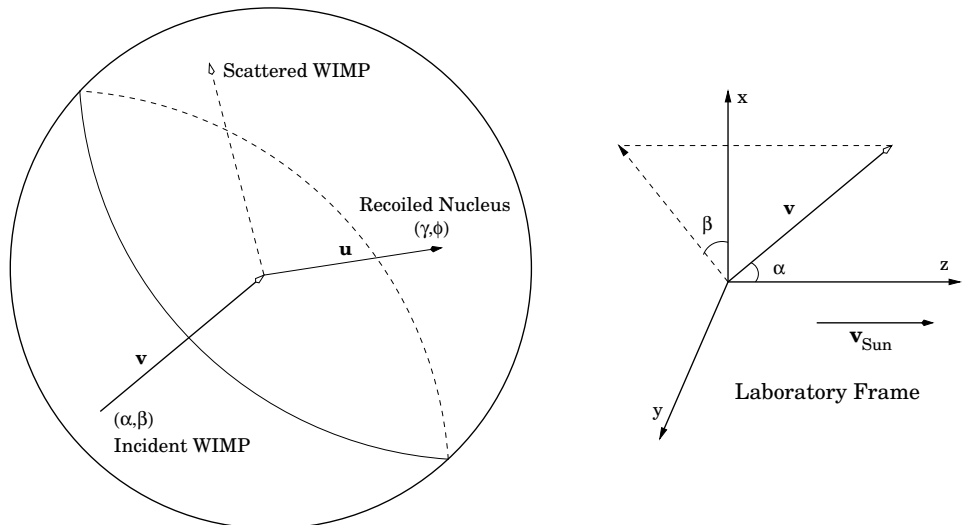,width=5in}}
  \caption{Geometry of a WIMP scattering event.  Here a WIMP is incident at an 
    angle $(\alpha,\beta)$ in the lab frame (shown on the right).  The WIMP
    hits a nucleus originally at rest that recoils at an angle $(\gamma,\phi)$ 
    in the lab frame.  The velocity of the Sun through the Galaxy defines
    the $z$-axis in the laboratory frame.}
  \label{fig:scattering}
\end{figure}

Consider a WIMP of mass $m_\chi$ moving with velocity $\vec v = v (\hat
x\sin\alpha \cos\beta + \hat y\sin\alpha\sin\beta + \hat z\cos\alpha)$ in the
laboratory frame (figure~\ref{fig:scattering}).  Let this WIMP scatter off
a nucleus of mass $m_n$ 
in the direction ($\theta^*$, $\xi$) in the center-of-mass frame.
The recoil energy of the nucleus is
\be Q = \frac{m_n m_\chi^2}{\left ( m_n + m_\chi \right)^2} v^2 (1-\mu), \ee
where $\mu = \cos\theta^*$.  Let $\vec u = u (\hat x \sin\gamma\cos\phi
 + \hat y\sin\gamma\sin\phi + \hat z\cos\gamma)$
be the recoil velocity of the nucleus in the laboratory frame
(figure~\ref{fig:scattering}).  To
determine the relations among these sets of angles
we begin by considering the simplified scattering problem of a WIMP incident
along the $z$-axis.  We can then rotate this result for an arbitrary WIMP
incident at an angle $(\alpha, \beta)$ to find (in the non-relativistic limit)
\bea
  \cos\gamma & = & \sqrt{\frac{1-\mu}2}\cos\alpha -
  \sqrt{\frac{1+\mu}2}\sin\alpha\cos\xi, \nonumber \\
  \sin\gamma\cos\phi & = & \sqrt{\frac{1-\mu}2}\sin\alpha\cos\beta -
  \sqrt{\frac{1+\mu}2}\left (\sin\beta\sin\xi - \cos\alpha\cos\beta\cos\xi
  \right), \\
  \sin\gamma\sin\phi & = & \sqrt{\frac{1-\mu}2}\sin\alpha\sin\beta +
  \sqrt{\frac{1+\mu}2}\left (\cos\beta\sin\xi + \cos\alpha\sin\beta\cos\xi
  \right). \nonumber
\eea
Notice in the limit of $(\alpha,\beta) \rightarrow (0,0)$ this reduces to the
usual two dimensional result.  

To remove the dependence on the center of mass
angles we note that 
\be 
  \sin \frac{\theta^*}2 = \sqrt{\frac{1-\mu}2} = \left| \cos\gamma\cos\alpha +
    \sin\gamma\sin\alpha \cos (\beta-\phi) \right|
\ee
and the Jacobian
\be
  \frac{\partial (\cos\gamma, \phi)}{\partial (\mu, \xi)} = \frac14
  \sqrt{\frac2{1-\mu}}.
\ee
Next, to simplify subsequent formulae, we define a geometrical factor, $J$, by
\begin{eqnarray}
  J (\alpha, \beta; \gamma, \phi) & \equiv &
  \frac{\vec{u}\cdot\vec{v}}{|\vec{u}| |\vec{v}|} = \frac14 \left|
    \frac{\partial (\cos\gamma, \phi)}{\partial (\mu, \xi)} \right|^{-1}
  \nonumber \\
  & = & \cos\alpha\cos\gamma + \sin\alpha\sin\gamma \cos (\beta-\phi).
  \label{eqn:J}
\end{eqnarray}
With this information we can now write the event rates solely as a
function of the incoming lab angle ($\alpha$, $\beta$) and the recoiling lab
angle ($\gamma$, $\phi$).

The event rate per nucleon in the detector is given by
\be 
  dR = f (v, \alpha, \beta) v^3 dv\, d (\cos\alpha)\, d\beta
  \frac{d\sigma}{d\mu}d\mu \frac{d\xi}{2\pi}.
\ee
Here we will assume the nucleon-WIMP scattering has the form
\be \frac{d\sigma}{d\mu} = \frac{\sigma_0}{2} F^2 (Q), \ee
where $F^2 (Q)$ is the form factor and $\sigma_0$ is energy independent.  The
number of events in which a nucleus
recoils with energy $Q$ is 
\be 
  \frac{dR}{dQ} = \frac{\sigma_0 \rho_0}{2m_r^2 m_n} F^2 (Q) \int_{v_{\rm
      min}}^{v_{\rm esc}} v\,dv \int d\Omega_{\alpha,\beta} f (v, \alpha,
  \beta),
\ee
where $v_{\rm esc}$ is the escape velocity of the Galaxy, $v_{\rm min}^2 =
\frac{\left(m_\chi + m_n \right)^2}{2m_\chi^2m_n}Q$ is the minimum incident
WIMP velocity that can produce a nuclear recoil of energy $Q$, and $m_r =
m_\chi m_n / (m_\chi + m_n)$ is the reduced mass.  This formula is well
known and, in the case of an isothermal model when we only consider the
projected motion of the Earth along the Sun's velocity, it can be evaluated 
analytically.

In this work we are most interested in the angular dependent event rate which
is found to be
\be
  \dRdOmega = \frac{\sigma_0\rho_0}{\pi m_n m_\chi} \int_{v_{\rm
      min}/J}^{v_{\rm esc}}v^3 dv\, F^2 \left(Q (v, J) \right)
  \int d\Omega_{\alpha,\beta} f (v,\alpha,\beta) J (\alpha,\beta; \gamma, \phi)
  \Theta (J).
\ee
Here $\Theta (J)$ is the usual step function and $J$ is the geometrical factor 
discussed above~(\ref{eqn:J}) that
relates the incident WIMP direction to the recoiled nucleus direction.

We can further derive the event rate as a function of both the
deposited energy and the recoil angle,
\be
  \frac{d^2R}{dQ\,d\Omega_{\gamma,\phi}} = \frac{m_n\sigma_0\rho_0}{8\pi
    m_r^4m_\chi} Q\, F^2 (Q) \int d\Omega_{\alpha,\beta}\, f\left ( v (Q,J),
    \alpha,\beta \right) \frac{\Theta (J)}{J^3}.
\ee
In principle this would provide the most information allowing for the best
separation of signal from background.  Note, however, that in order to
fully exploit this signature some detector-specific estimation of the
energy dependence of various backgrounds must be given.  

In this work we will focus on the angular event rate, which does not
require a knowledge of the specific energy dependent detector backgrounds
in order to explore the separation of signals from backgrounds. 
Exploiting the full distribution for various proposed detectors will be
analyzed in future work.

\section{Model Parameters}

Armed with the above differential event rates we can now analyze a wide
variety of halo models, form factors, WIMP parameters, detector parameters, 
{\it etc}.  Here we will discuss the parameters considered in this study.

\subsection{Halo Models}

A standard spherically symmetric isothermal sphere reproduces the large
scale features of the flat rotation curves around galaxies.   However,
numerous dynamical arguments suggest that actual halo model may not be
well described by such a distribution.  As a result, a variety of more
complex analytical models have been developed.  
The models we discuss here can be
considered to be of three types: smooth axisymmetric, smooth triaxial, and
clumped.  The first type includes a standard spherical isothermal WIMP halo, as
well as a modified axisymmetric halo, both rotating and non-rotating.  
Most recently a general triaxial generalization
of the spherical isothermal halo has been developed.  As an example of
the last type of halo model, which ultimately should include realistic
halo distributions arising from accretion of sub-systems by the growing
galaxy, we consider a caustic model recently developed by Sikivie.
While all of these models are analytic approximations, they do represent
a broad spectrum of different possibilities, which should give some idea
of the range of uncertainty in the predicted angular distributions in
detectors.  

\subsubsection{Isothermal Model}

A simple spherically symmetric isothermal distribution is given by,
\be 
  f (\vec v) = \frac1{\pi^{3/2}v_0^3} e^{-\left.{\left|\vec v\right|^2}\right/
    {v_0^2}}.
\ee
Here $\left<v^2\right>^{1/2} = \sqrt{3/2}\,v_0$ is the dispersion velocity of
the dark matter in the halo.  Note that we can allow different dispersions in
each direction but will not pursue that option further.  The standard value
chosen for $v_0$ is $220\kmps$.  We will consider the range of values
$v_0=150\kmps$, $220\kmps$, and $300\kmps$ here.

\subsubsection{Axisymmetric Halo Model}

The Evans model~\cite{evans} is an axisymmetric halo model that allows for
flattening.  This model has been studied in the context of annular modulations 
by Kamionkowski and Kinkhabwala~\cite{kamionkowski}.  The distribution
function is given by
\be
  f (\vec v) = \frac1{D^2} \left[ AR_0 \left (v\cos\alpha - \vex\right)^2 +
    B\right] e^{-2\left.{(v-\Psi)^2}\right/{v_0^2}}\;
  e^{-2\left.{v_\oplus^2-\Psi^2}\right/{v_0^2}} + \frac CD
  e^{-\left.{(v-\Psi)^2}\right/{v_0^2}}\;
  e^{-\left.{v_\oplus^2-\Psi^2}\right/{v_0^2}},
\ee
where
$$
  A = \left ( \frac2\pi \right)^{5/2} \frac{1-q^2}{Gq^2v_0^3}, \qquad
  B = \sqrt{\frac2{\pi^5}} \frac{R_c^2}{Gq^2v_0}, \qquad
  C = \frac{2q^2 - 1}{4\pi Gq^2v_0},
$$
\be
  D = R_c^2+R_0^2, \qquad
  \Psi = \vex \cos\alpha + \vey\sin\alpha\cos\beta
  + \vez\sin\alpha\sin\beta,
\ee
$R_c$ is the core radius, $R_0$ is the distance of our solar
system from the center of the Galaxy, $v_0$ is the circular speed at large
radii, and $q$ is the flattening parameter ranging from $q=1$ to
$q=1/\sqrt2$.  In this work we adopt $R_c=7\kpc$, $R_0=8.5\kpc$, and
$v_0=220\kmps$.  The most important parameter is the flattening, $q$.  Here we 
choose $q=1$ (cored isothermal model), $0.85$, and $1/\sqrt2$ (maximum
flattening).

\subsubsection{Triaxial Halo Models}

Constructing velocity distributions from general triaxial halos is a
difficult problem.  Evans, Carollo, and de Zeeuw~\cite{triaxial} have
provided such distributions for the logarithmic ellipsoidal potential; the
simplest triaxial model and a natural generalization of the isothermal
sphere.  In this case we can approximate the velocity distribution as 
\be
  f (\vec v) = \frac1{\pi^{3/2}\sigma_x\sigma_y\sigma_z}\exp \left[
    -\frac{v_x^2}{\sigma_x^2} -\frac{v_y^2}{\sigma_y^2} -
    \frac{v_z^2}{\sigma_z^2} \right].
\ee
When the Sun is on the long axis of the ellipsoid we have the relations
\be
  \sigma_x^2 = \frac{v_0^2}{(2+\gamma) (p^{-2}+q^{-2}-1)}, \quad
  \sigma_y^2 = \frac{v_0^2 (2p^{-2}-1)}{2(p^{-2}+q^{-2}-1)}, \quad
  \sigma_z^2 = \frac{v_0^2 (2q^{-2}-1)}{2(p^{-2}+q^{-2}-1)}.
\ee
When the Sun is on the intermediate axis of the ellipsoid we have the relations
\be
  \sigma_x^2 = \frac{v_0^2 p^{-4}}{(2+\gamma) (1+q^{-2}-p^{-2})}, \quad
  \sigma_y^2 = \frac{v_0^2 (2-p^{-2})}{2(1+q^{-2}-p^{-2})}, \quad
  \sigma_z^2 = \frac{v_0^2 (2q^{-2}-p^{-2})}{2(1+q^{-2}-p^{-2})}.
\ee
Here $p$ and $q$ describe the axis ratios of the ellipsoid ($p=q=1$ is a
spheroid) and $\gamma$ describes the anisotropy ($\gamma=1$ for a sphere).
We will consider the cases where $p=0.9$ and $q=0.8$ for both positions of
the Sun.  We will also consider $\gamma=-1.78$ (radially anisotropic) and
$\gamma=16$ (tangentially anisotropic).

\subsubsection{Galactic Infall---Sikivie Caustic Model}

The caustic model is based on the work of Sikivie and collaborators
\cite{caustic}.  It is derived from the assumption that WIMPs continuously fall
into our Galaxy from all directions.  The WIMPs only get thermalized after
many passes through the Galaxy.  Thus the WIMP halo of our Galaxy should be
made up of an isothermal (or similar) core plus a set of inflowing and
outgoing peaks in velocity space with the distribution function given by
\be
  f (\vec v) = \sum_j \rho_j \delta (\vec v - \vec v_j).
\ee
Here $j$ sums over the velocity flows, $\rho_j$ is the density in each flow,
and $\vec v_j$ is the velocity of each flow.
The peaks for one such model are given in
table~\ref{tab:caustic}.  Notice that there are two flows for each velocity
peak.
In this model the total local density of WIMPs is $\rho_0 =
0.52\unit{GeV/cm^3}$ and the first 14 peaks are not thermalized (Sikivie,
private communications).  From table~\ref{tab:caustic} we find $\rho_{\rm
  caustic} = 0.34\unit{GeV/cm^3}$.  Thus 65\% of the WIMPs are in the form of
velocity flows and the remaining 35\% are in the form of a thermalized
distribution.  For simplicity we will choose the standard ($v_0=220\kmps$)
isothermal model for the thermalized distribution.

Most recently a number of groups have begun to numerically investigate the
growth of our Galactic halo via accretion in N-body simulations.  While we 
plan to
incorporate these results into a future analysis, it is worth stressing that
while the Sikivie model is undoubtedly an idealization of the actual
hierarchical accretion process, the resulting distribution, containing a 
thermalized background plus several different unthermalized velocity flows
may share some qualitative features of the N-body distribution, even
if it is not likely to agree in detail with these results.  It thus
provides one tractable example of the complexity that might actually
characterize the real galactic halo, and can usefully demonstrate 
some of the possible implications of phase
space flows for signatures in WIMP detectors.

\subsection{Motion of the Earth}

In previous analyses it was generally assumed that the Earth orbits the Sun in the
same plane that the Sun orbits the center of the Galaxy.  In fact this is not true,
the Earth's orbit is tilted by approximately $60^\circ$ from this plane. 
When working with an isothermal model and considering either the total number of
events or the annular modulation, the effect of the tilt of the Earth's orbit is
insignificant.  However, since we are interested here in angular
effects and in halo models that need not be spherically symmetric, it
is necessary to use the correct velocity of the Earth.  In
the frame of the Galaxy this is 
\bea
  \vex & = & 0.13 v_\odot \sin \left(2\pi (t-t_p) / {\rm year}\right), \nonumber \\
  \vey & = & -0.11 v_\odot \cos \left(2\pi (t-t_p) / {\rm year}\right), \\
  \vez & = & v_\odot \left[ 1.05 + 0.06 \cos \left(2\pi (t-t_p) / {\rm year}\right)\right],
  \nonumber
\eea
where $v_\odot = 220\kmps$ is the velocity of the Sun in the frame of the
Galaxy, $t$ is measured in days, and $t_p \approx 153\unit{days}$ (June 2) is
the day when the direction of the Earth's motion around the Sun matches the
direction of the Sun's motion around the Galactic center.  The numerical
coefficients reflect the $60^\circ$ tilt of the Earth's orbit.

The distribution functions above are given in the rest frame of the Galaxy.
In the lab frame (rest frame of the Earth) we must shift them as $f (\vec v +
\vec v_\oplus)$.

\subsection{Form Factor}

In WIMP-nucleus scattering, a form factor is incorporated into the cross section, as
described above, to account for the loss of coherence over the nucleus for large
momentum transfers.   The appropriate specific form factor is, of course,
nucleus-dependent.  However, for the purposes of this analyses we utilize several
reasonable analytical approximations which are appropriate in the case of either
spin-dependent, or spin-independent scattering, as described below. 
For a more detailed discussion  see~\cite{review}.

\subsubsection{Spin Independent}

Here the WIMP couples to various quantum numbers of the entire nucleus.  The standard
form factor,
$F^2 (Q)$, for such nuclear interactions is the Wood-Saxon  form factor.  In this work
we will consider the simpler exponential form factor
\be F^2 (Q) = e^{-Q/Q_0} \ee
since it is easier to work with analytically and it produces qualitatively
similar results.  Here $Q_0 = {3}\left/ (2m_n r_0^2)\right.$ and
$r_0=0.3+0.91\sqrt[3]{m_n}$ is the radius of the nucleus (in femtometers when
$m_n$ is in GeV).

\subsubsection{Spin Dependent}

The energy dependent component of the form factor for an axial vector
interaction is given by 
\be 
  F^2 (q) \propto S (q) = a_0^2 S_{00} (q) + a_1^2 S_{11} (q) + a_0 a_1
S_{01} (q).
\ee
Here $a_0$ ($a_1$) represents the isoscalar (isovector) parameterization of 
the matrix element.  $S_{00}$, $S_{11}$, and $S_{01}$ are obtained from
detailed nuclear calculations.  In this work we will parameterize this
energy dependent part using exponential functions for analytical ease.

For the purposes of our calculation we choose germanium.  While it is not
likely that this material will in fact be used in direction-sensitive detectors, it
has the virtue of being a mid-range nucleus with sensitivity to 
axial couplings, and has been well studied.  Without specific candidates for
directional detector targets it thus seems a reasonable first step, in order to get
some idea of the possible changes in signature for spin-dependent vs.\ spin
independent interactions.   For germanium, we find
\be
  S_{\rm Ge} (q) = 0.20313 \left ( 1.102 a_0^2 e^{-7.468y} + a_1^2 e^{-8.856y}
- 2.099 a_0 a_1 e^{-8.191y} \right),
\ee
where $y = \left ( (1\unit{fm}) |\vec q|/2 A^{1/6}\right)^2$ and is valid for 
$y < 0.2$~\cite{Gespin}.  For a WIMP that is a pure bino the couplings are
$a_0 = 0.082 \zeta_q$ and $a_1 = 0.276 \zeta_q$ when the European Muon
Collaboration values for the spin content of the nucleons are
employed.  For our purposes in this work $\zeta_q$ is an energy independent, 
overall scale factor.  See~\cite{Gespin} for a more thorough discussion of
$S_{\rm Ge}$.

\subsection{Rotating Halo}

To allow for a rotating halo we employ the standard
technique from Lynden-Bell~\cite{lynden-bell}.  Let
\be 
  f_+ (v, \alpha, \beta) = \left\{ 
    \begin{array}{ll}
      f (v, \alpha, \beta) & 0 < \alpha < \pi/2 \\
      0 & \pi/2 < \alpha < \pi
    \end{array}
    \right.
\ee
and
\be 
  f_- (v, \alpha, \beta) = \left\{ 
    \begin{array}{ll}
      0 & 0 < \alpha < \pi/2 \\
      f (v, \alpha, \beta) & \pi/2 < \alpha < \pi
    \end{array}
    \right. .
\ee
Then the rotating distribution function is given by
\be
  f_R (v, \alpha, \beta) = (1+\kappa) f_+ (v,\alpha,\beta) 
  + (1-\kappa) f_- (v,\alpha,\beta).
\ee
Note that this definition does not change the normalization nor dispersion of
the distribution function; $\bigl<f_R\bigr> = \bigl<f\bigr>$ and $\bigl<v^2
f_R\bigr> = \bigl<v^2 f\bigr>$, independent of $\kappa$.  When $\kappa = 0$ we 
recover the original distribution function.

To generate a halo with a particular average velocity we note that 
\be 
  \left< v f_R \right> = \frac{2v_0}{\sqrt\pi} \kappa.
\ee
Thus when $\left< v f_R \right> = v_0$ we find that $\kappa = \sqrt\pi / 2$.

\subsection{Target Nucleus}

As mentioned above we chose \Ge\ as our prototypical nuclear target.  Germanium has
both scalar and axial vector interactions with WIMPs, and thus serves as a
good baseline for probing the dependencies in the angular distribution.
Throughout, we assume that 25\% of the recoil energy is in the detected
channel (quenching factor of $0.25$).  The detector thresholds quoted below
have taken this into account.
Both argon and xenon have been discusses as likely targets in a TPC such as 
DRIFT~\cite{tpc}.  There are experimental difficulties with using argon
(it is naturally radioactive) and little is known about xenon gas (in
particular its form factor).  Thus we will not examine these targets in detail
at this time.  

Note that for our purposes the choice of \Ge\ as target is somewhat arbitrary. 
Since the normalization of the overall  WIMP-nucleus scattering cross
section is not relevant for our discussions, the chief distinction between
choosing different target nuclei will be the kinematic  dependence on WIMP mass vs
target mass, and possible alterations in the  spin-dependent form factor
parameters.   For most materials being used for  dark matter detection, quenching
factors generally lie in the $0.1-0.4$ range and thus our choice of the measured Ge
quenching factor is also relatively generic.

\subsection{WIMP parameters}

Throughout this work we focus on the shape of the nuclear recoil
distribution.  Thus, as emphasized above, we do not focus on the overall event rate
normalization.  This quantity depends on specifics of the halo density, the WIMP mass
and couplings, the
target nucleus, detector size, type of interaction, {\it etc}.  The key
point is that when modeling a specific detector this normalization will
be essential, but in order to explore generic features of the angular
signature of WIMP-nucleus scattering, it is not important. 

        As noted above, the mass of the WIMP affects both the normalization and shape
of the nuclear recoil distribution.  Thus, the results remain identical if one
appropriate scales both target nucleus and WIMP mass.  In this work
we will consider
$m_\chi=60\GeV$ as our standard WIMP mass (as mass matches well with the
germanium target, $m_{\rm Ge}=73\GeV$).  In some cases we will also show
results for $m_\chi=180\GeV$ to show how the mass of the WIMP affects the
results.  The results we present should thus be considered to be appropriate
to a WIMP whose mass matches well with the target nucleus mass, or in some cases
greatly exceeds it. 

\section{Results}

The calculations of $dR/d\Omega$ were performed on a $40\times40$ grid in
the $(\cos\gamma,\phi)$ plane.  They were calculated in $5$ day bins and
summed over the year.  By summing over the year we average out any annual
modulations in both the signal and the background.  Furthermore, we are
implicitly averaging over the day in which time the detector rotates (along 
with the Earth) by $2\pi$ with respect to the direction of motion.  Thus
any angular dependent backgrounds, due to, for example, hot spots in the detector,
will also be averaged out (at least for the component in the plane of motion of the
Earth) and the probability of a background induced nuclear
recoil is uniform in angle when averaged over the year, with respect to the direction
of the Earth's motion around the Sun.

\subsection{Angular Distributions}

\begin{figure}[tbp]
  \leavevmode\center{\epsfig{figure=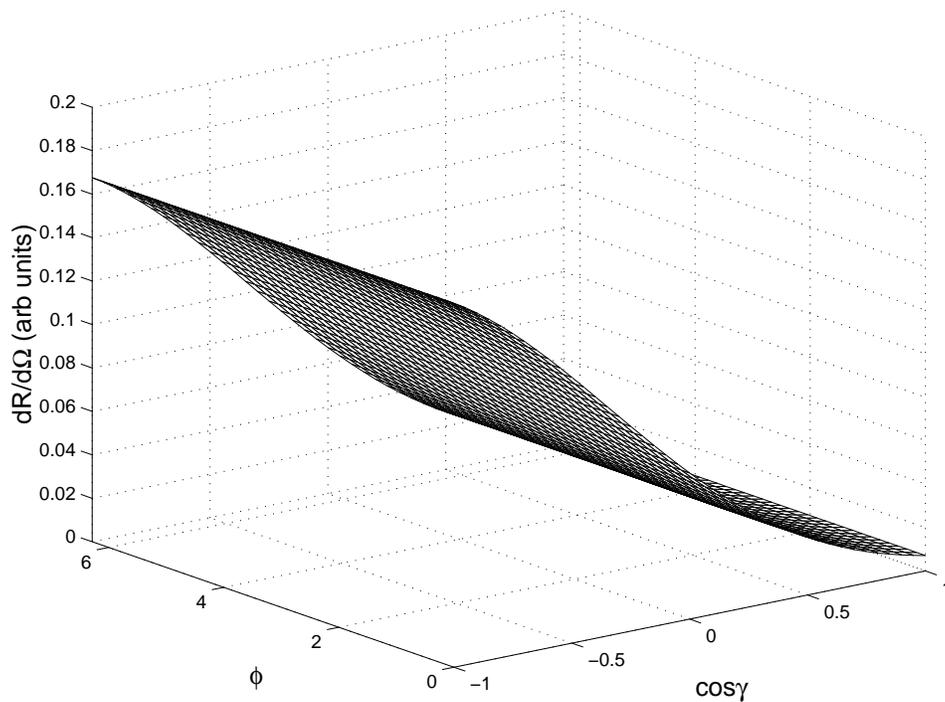,width=5in}}
  \caption{The angular distribution of nuclear recoil events, $dR/d\Omega$ for 
    an isothermal halo model.  Here $v_0=220\kmps$ and $\Eth=0\keV$.}
  \label{fig:angular-iso220}
\end{figure}

\begin{figure}[tbp]
  \leavevmode\hbox to\hsize
  {\hss\epsfig{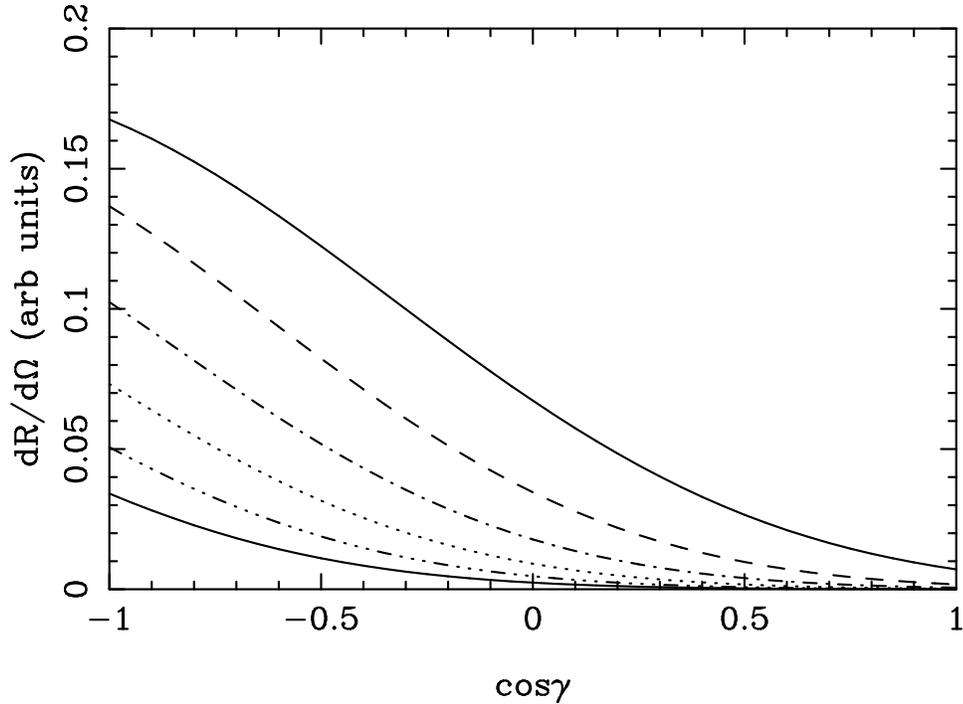}\hss}
  \caption{The angular distribution of nuclear recoil events, $dR/d\Omega$ for 
    an isothermal halo model with $\phi=0$ and $v_0=220\kmps$ as a function of
    $\cos\gamma$ and for various threshold energies.  Here the thresholds are
    (from upper to lower) $\Eth = 0\keV$ (upper solid curve), $2\keV$, $4\keV$, 
    $6\keV$, $8\keV$, and $10\keV$ (lower solid curve).}
  \label{fig:iso220-gamma}
\end{figure}

The angular distribution for the isothermal model with $v_0=220\kmps$ is shown 
in figure~\ref{fig:angular-iso220}.  Notice the exponential decay in
$\cos\gamma$ in the forward direction and the independence of $\phi$.  The reason
for the decay in the forward direction is simple.  The Earth is moving through 
this WIMP halo, and thus more WIMPs are incident from the forward direction than
the backward direction, and thus nuclei are preferentially scattered backwards.  This
exponential decay would lead
to an easy statistical separation of signal events from (flat) background
events as discussed below.  To examine dependence on threshold we plot
$dR/d\Omega$ for $\phi=0$ for a number of
different threshold energies (figure~\ref{fig:iso220-gamma}).  As we increase
the threshold the backward
scattering peak in the distribution becomes more pronounced in relation to
the rest of the distribution, making
it more easily distinguishable from a (flat) background.  However,
as is also evident from the figure, as we increase the threshold the total
number of events decreases.  These two effects will be discussed below.

\begin{figure}[tbp]
  \leavevmode\hbox to\hsize
  {\hss\epsfig{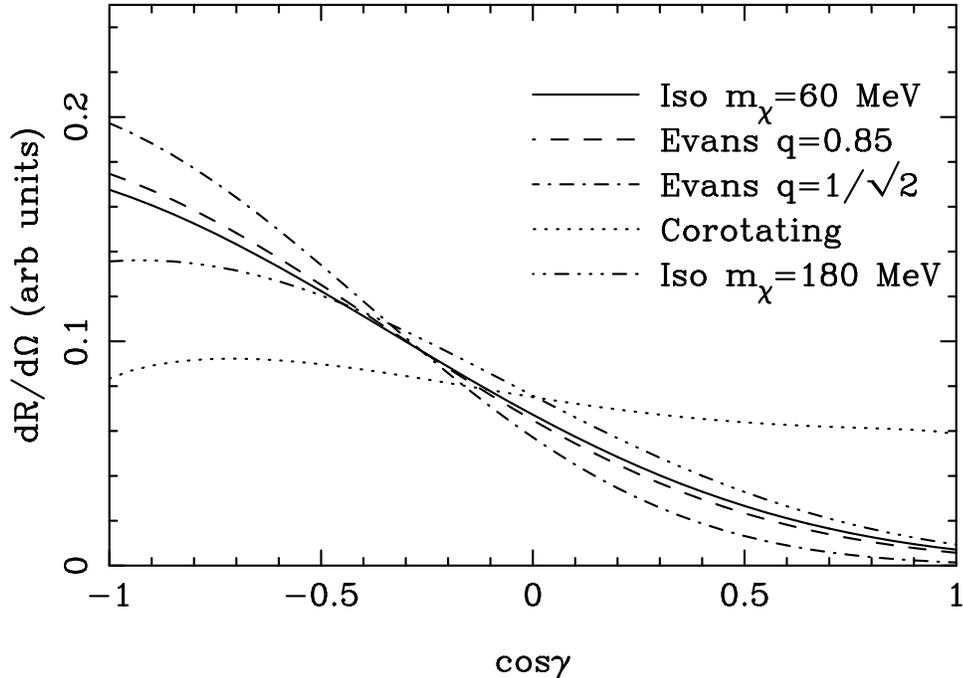}\hss}
  \caption{The angular distribution of nuclear recoil events, $dR/d\Omega$ for 
    the isothermal and Evans halo models with $\phi=0$ $\Eth=0\keV$ as a
    function of $\cos\gamma$.  The Evans model with $q=1$ is not shown
    since it is nearly identical to the isothermal model with
    $v_0=220\kmps$.  There are two curves shown for the $v_0=220\kmps$
    isothermal model; one for $m_\chi=60\GeV$ (solid line) and the other
    for $m_\chi=180\GeV$ (dashed-dotted line).  The Evans model with
    $q=0.85$ (dashed line) and $q=1/\sqrt2$ (long dashed-short dashed line)
    are also shown.  Finally the corotating model (dotted line) is shown.
    See the text for more details.}
  \label{fig:models-gamma}
\end{figure}

The angular distributions for a number of isothermal, axisymmetric and 
rotating models are
shown in figure~\ref{fig:models-gamma}.  Here we plot $dR/d\Omega$ as a function of
$\cos\gamma$ for $\phi=0$ and $\Eth=0\keV$.  The two solid curves for the
$v_0=220\kmps$ isothermal model are for two different WIMP masses,
$m_\chi=60\GeV$ and $m_\chi=180\GeV$.  All of the curves are normalized so
that $R=1$.  As can be seen from the figure all of the models are similar
in shape (at least at $\Eth=0\keV$).  Thus we do not expect a large
variation in the number of events required to identify a WIMP signal as we
will discuss later.  On the other hand, it also means that it will be
difficult to distinguish between different halo models based solely on this 
information.

\begin{figure}[tbp]
  \leavevmode\hbox to\hsize{\epsfig{figure=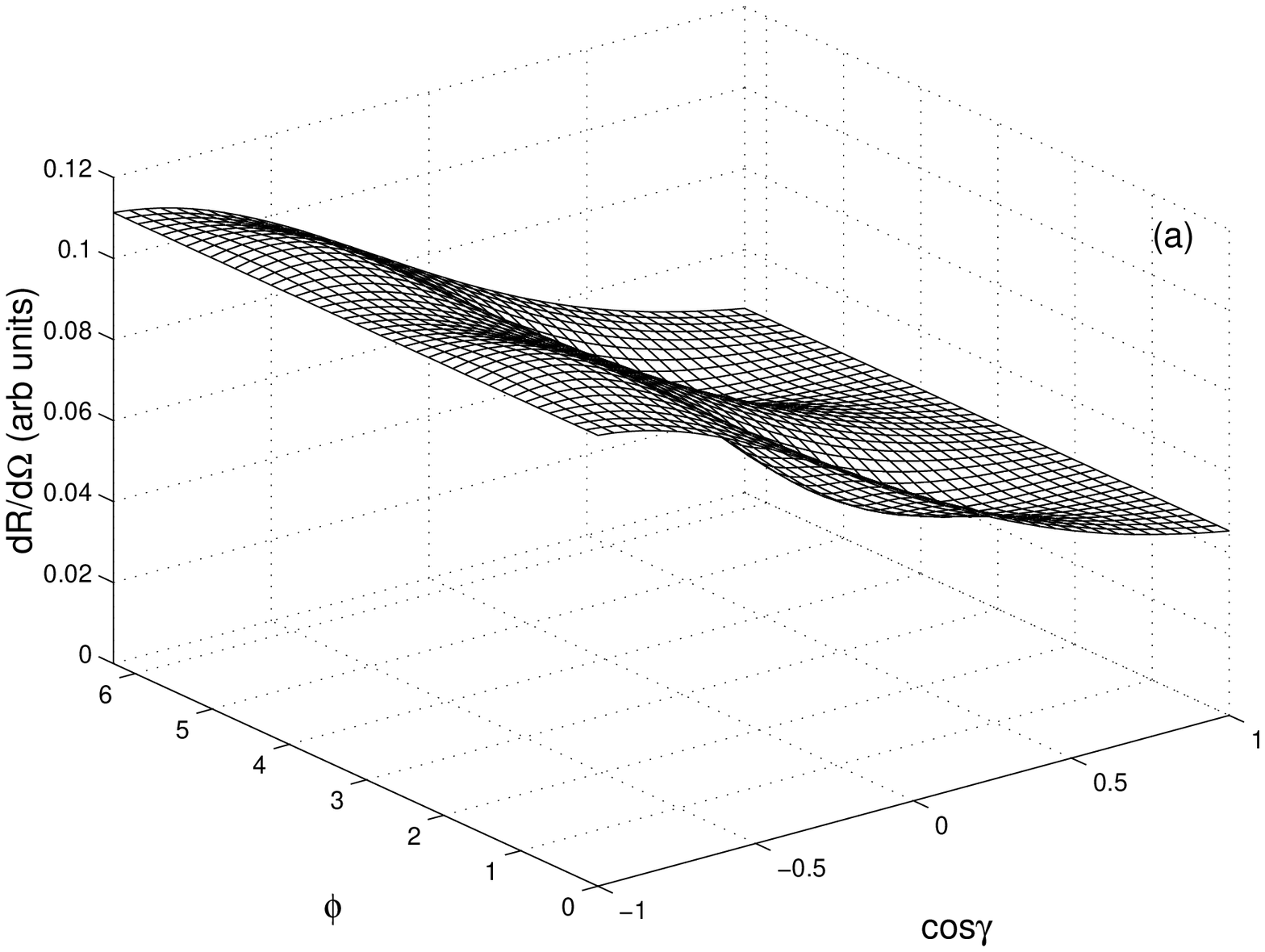,width=3.5in}\hfil
    \epsfig{figure=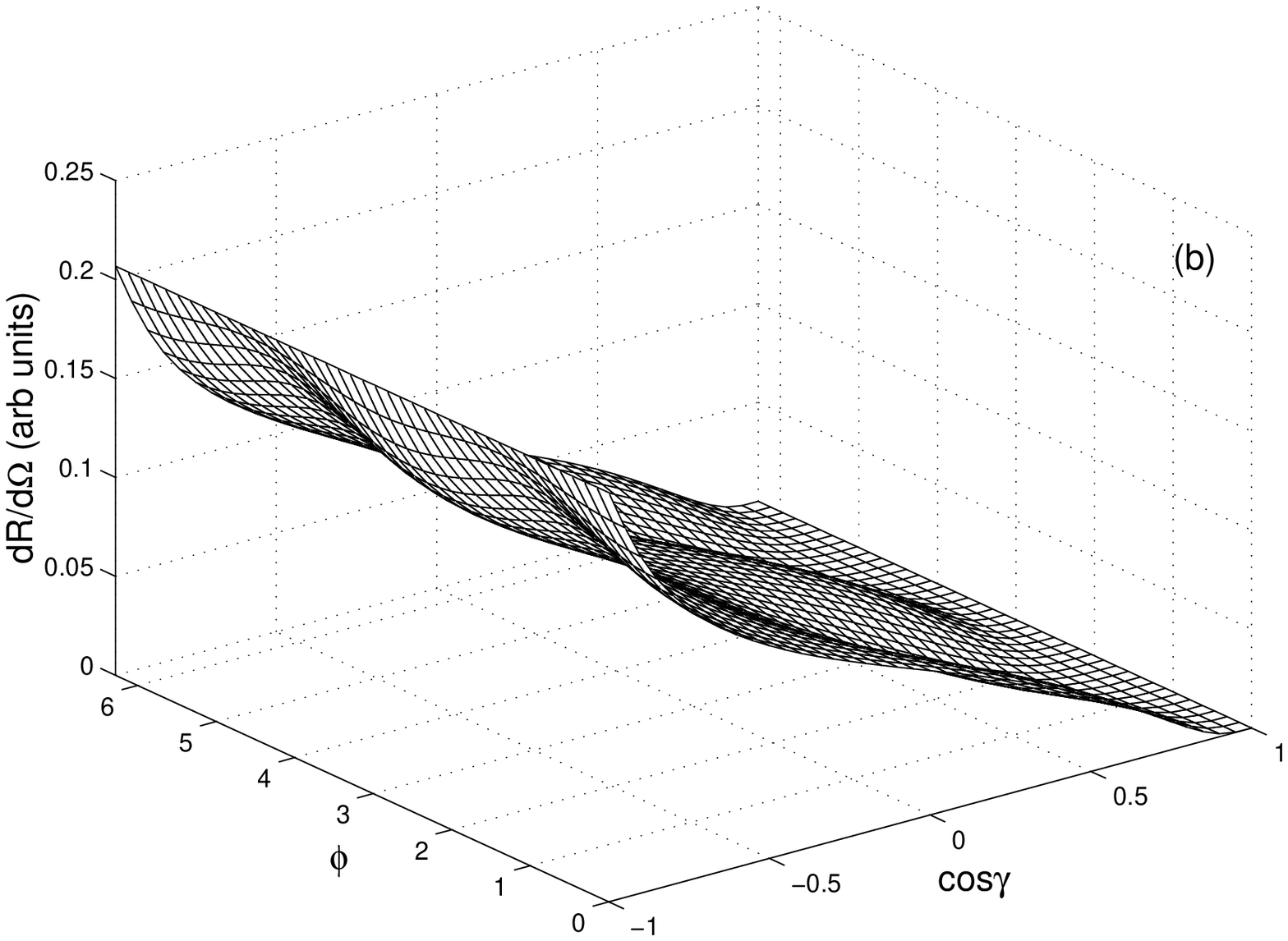,width=3.5in}}
  \leavevmode\hbox to\hsize{\epsfig{figure=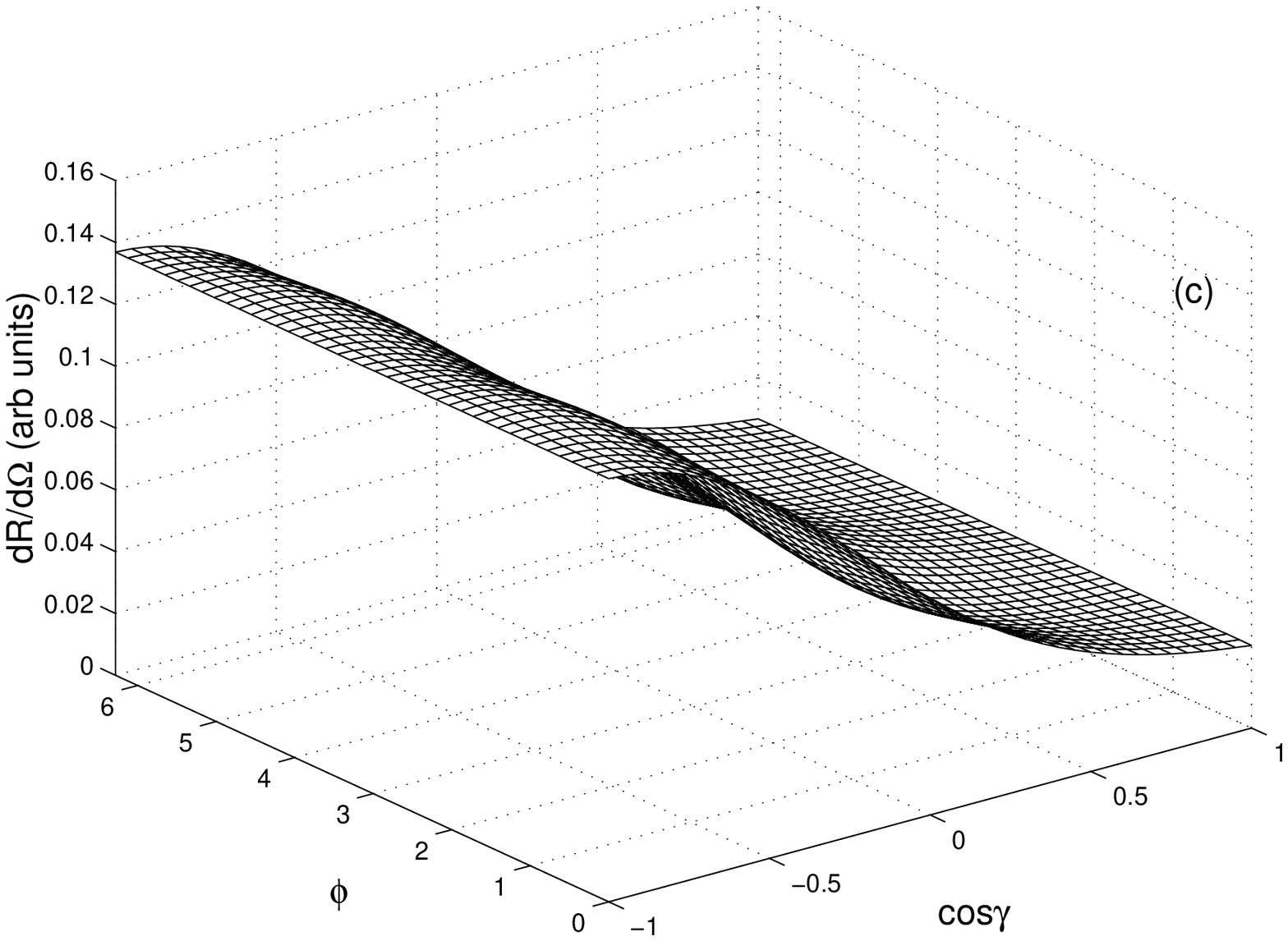,width=3.5in}\hfil
    \epsfig{figure=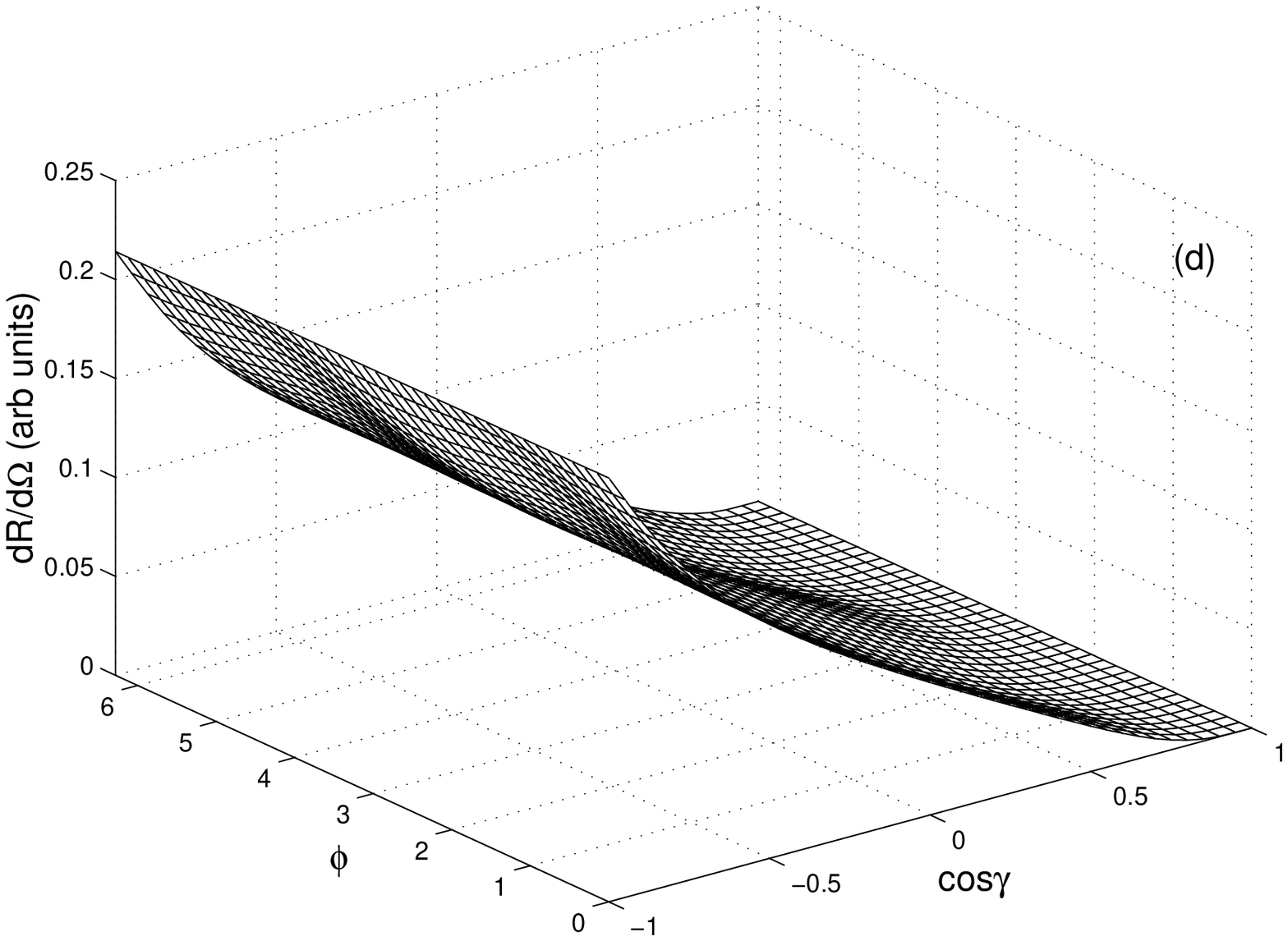,width=3.5in}}
  \caption{The angular distribution of nuclear recoil events, $dR/d\Omega$ for 
    the triaxial model for $\Eth=0\keV$.  In all cases $p=0.9$ and
    $q=0.8$.  We consider the Sun on the intermediate axis (a)
    $\gamma=-1.78$ (radially anisotropic) and (b) $\gamma=16$ (tangentially
    anisotropic). We also consider the Sun on the major axis (c)
    $\gamma=-1.78$ (radially anisotropic) and (d) $\gamma=16$ (tangentially
    anisotropic).}
  \label{fig:angular-tri}
\end{figure}

The angular distribution for the triaxial model is shown in
figure~\ref{fig:angular-tri}.  Here there is a panel for each set of
parameters.  This shows how much the rate can vary even in a single model
depending on the parameters.  The most difficult combination of parameters to
distinguish from background is seen in panel~(a) since it looks the most
similar to a flat distribution.  The rest of the models
look more like the isothermal distribution (panels~b--d) and will be
easier to distinguish from the background.

\begin{figure}[tbp]
  \leavevmode\center{\epsfig{figure=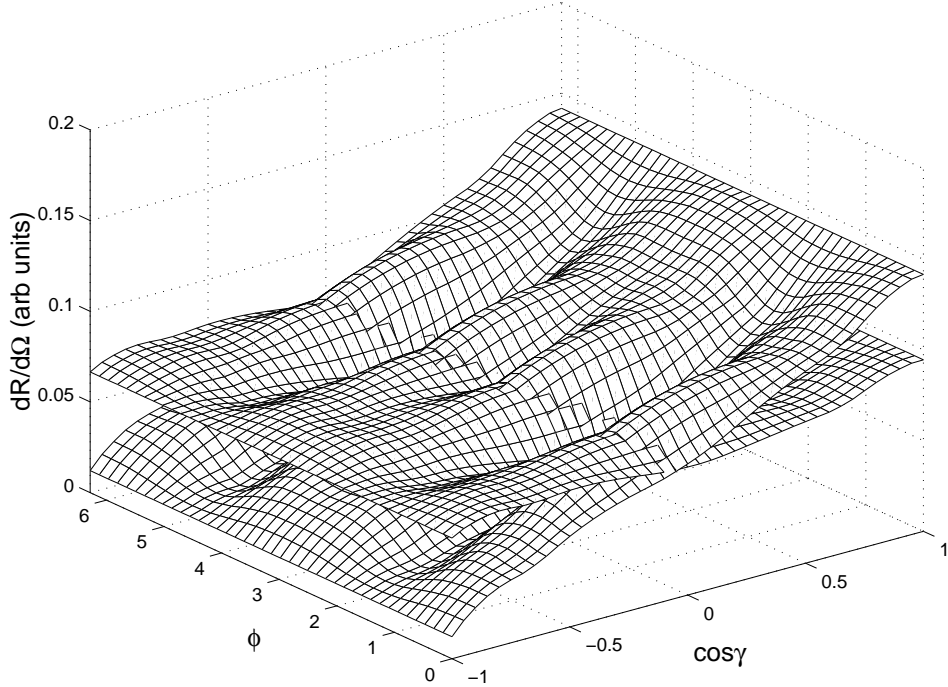,width=5in}}
  \caption{The angular distribution of nuclear recoil events, $dR/d\Omega$ for 
    the caustic model.  Here the contribution from the caustics (steeper) and
    the actual caustic plus $v_0=220\kmps$ isothermal (flatter) distributions
    are shown for $\Eth=0\keV$.}
  \label{fig:angular-caustic}
\end{figure}

The angular distribution for the caustic model is shown in
figure~\ref{fig:angular-caustic}.  Here both the pure caustic and a caustic with
isothermal component are shown.  Notice that the pure caustic model is peaked
in the forward direction (opposite the other models) and falls off more
slowly.  This is because we are located near a caustic and WIMPS
are preferentially catching up with the Sun as it moves through the Galaxy, so more
are incident from behind than from the forward direction.
There are also a number of small features in the
$\phi$ direction. When combined with the $v_0=220\kmps$ isothermal model the
distribution flattens greatly.  This makes is more difficult to distinguish from a
flat background than the other models.

\subsection{Signal Identification}

To determine the number of signal events necessary to identify a WIMP signal
we employ a maximum likelihood analysis along with Monte Carlo generation of
sample nuclear recoil distributions.  We define a likelihood function
\be {\cal L} = \prod_{i=1}^{N_e} P (\cos\gamma_i, \phi_i), \ee
where $N_e$ is the total number of events and $P (\cos\gamma_i, \phi_i)$ is
the probability of nuclear recoil in the $(\gamma_i, \phi_i)$ direction for a
particular model (e.g.\ an isothermal model).  
Here $P (\cos\gamma_i, \phi_i)$ is generated by calculating $dR/d\Omega$ for
each recoil direction $(\gamma_i, \phi_i)$ and averaging the result over the
year in 5 day bins.  At the 95\% confidence level
when $\log {\cal L}_{dR/d\Omega} - \log {\cal L}_{\rm flat} < 1$ the
distributions
are indistinguishable.  We generate 10,000 data sets for each $N_e$ and demand 
that the log-likelihood condition is satisfied less than 5\% of the time.  The
smallest $N_e$ for which this occurs is the minimum number of events required
to get a 95\% detection 95\% of the time.

With this procedure we can determine the number of events required to
distinguish a signal from a flat background or to distinguish one signal
(e.g.\ an isothermal halo) from another signal (e.g.\ a caustic halo).  For
a pure signal we use $P_{dR/d\Omega}$ as determined by the shape of
$dR/d\Omega$.  For a signal-to-noise ratio, $S/N$, we replace $P$ with
$P_{\rm total} = \lambda P_{\rm flat} + (1-\lambda) P_{dR/d\Omega}$ where
$\lambda = 1/ (1+S/N)$.  For the results discussed here we consider a
$40\times 40$ grid in the $(\cos\gamma,\phi)$ plane and assume perfect
angular knowledge on this grid.  This implies an accurate angular recoil
detector.  We also consider the case where knowledge of just the
forward-backward asymmetry is available.  In this case $P$ is replaced by a
binomial distribution.

\begin{figure}[tbp]
  \leavevmode\hbox to\hsize
  {\hss\epsfig{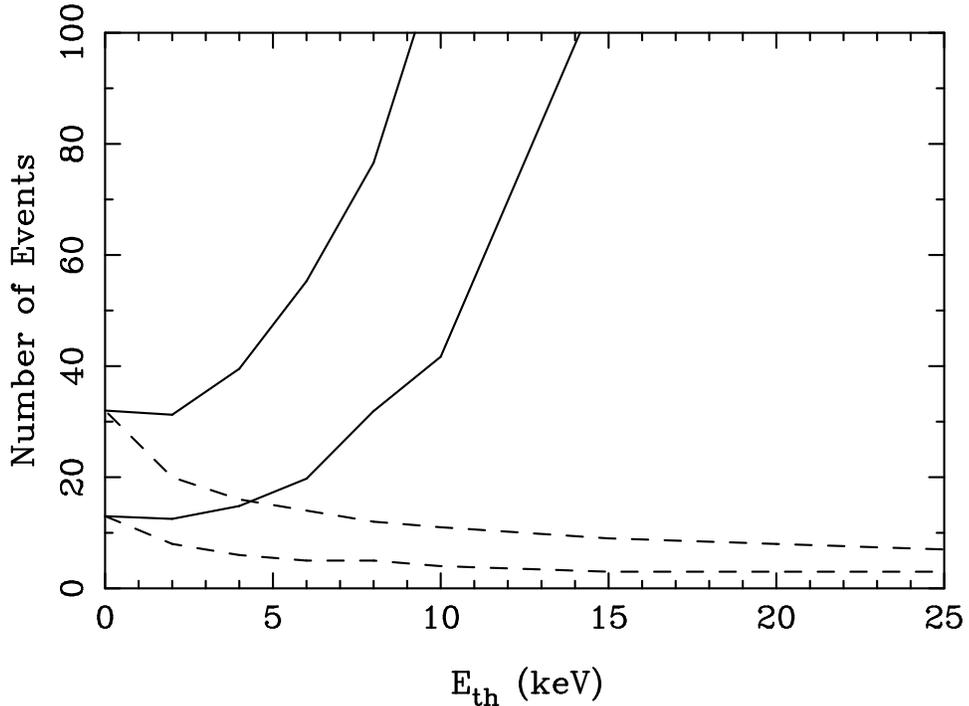}\hss}
  \caption{The number of signal events required as a function of threshold
    energy to distinguish an isothermal distribution of WIMPs in the halo
    with $v_0=220\kmps$ from a flat background.  The bottom set of curves
    is for a pure signal and the top set of curves is for a signal-to-noise
    ratio of 1.  The dashed curves show the number of events needed as a
    function of threshold energy.  The solid curves show the number of
    events needed for $\Eth=0\keV$ in order to achieve the necessary number
of events at the higher thresholds. 
These curves include the fact
    that the number of events detected falls quickly as the threshold is
    increased.  See the text for more details.}
  \label{fig:Nevents-iso220}
\end{figure}

In figure~\ref{fig:Nevents-iso220} we show the number of signal events
needed to differentiate a WIMP signal from a flat background at the 95\%
confidence level 95\% of the time for a standard isothermal
halo.  In this figure there are
two sets of curves shown.  The lower set contains the results for a pure signal (no
noise).  The upper set contains the results for a signal to noise ratio of
one, thus twice as many events as shown would have been detected (roughly
half signal and half noise).  In each set two curves are shown.  The lower
curve in each set shows the actual number of events required to distinguish
the WIMP signal.  For the S/N=1 case this decreases from 32 events at
$\Eth=0\keV$ to 8 events at $\Eth=25\keV$.  As mentioned above, this is due
to the fact that the distribution becomes more peaked in the backward
direction, thus it is more easily distinguished from a flat distribution.
However, as also mentioned above, at a higher threshold energy there are
fewer events.  Thus, for a fixed-size detector raising the threshold will
not in general aid in signal detection, because most of the signal events
will be discarded.   We attempt to account for this fact in
the upper curve in each set.  The upper curve is gives the number of events that would
have been detected at zero threshold in order to observe the number of events shown
in the lower curve for finite thresholds.  This curve shows the effect of the decrease
in the rate as the threshold is increased.  For isothermal halos, at least, it is
desirable to choose the lowest possible threshold of the detector for a fixed
signal-to-noise ratio in order to improve the ability to detect WIMPs with the fewest
number of events.

\begin{figure}[tbp]
  \leavevmode\hbox to\hsize
  {\hss\epsfig{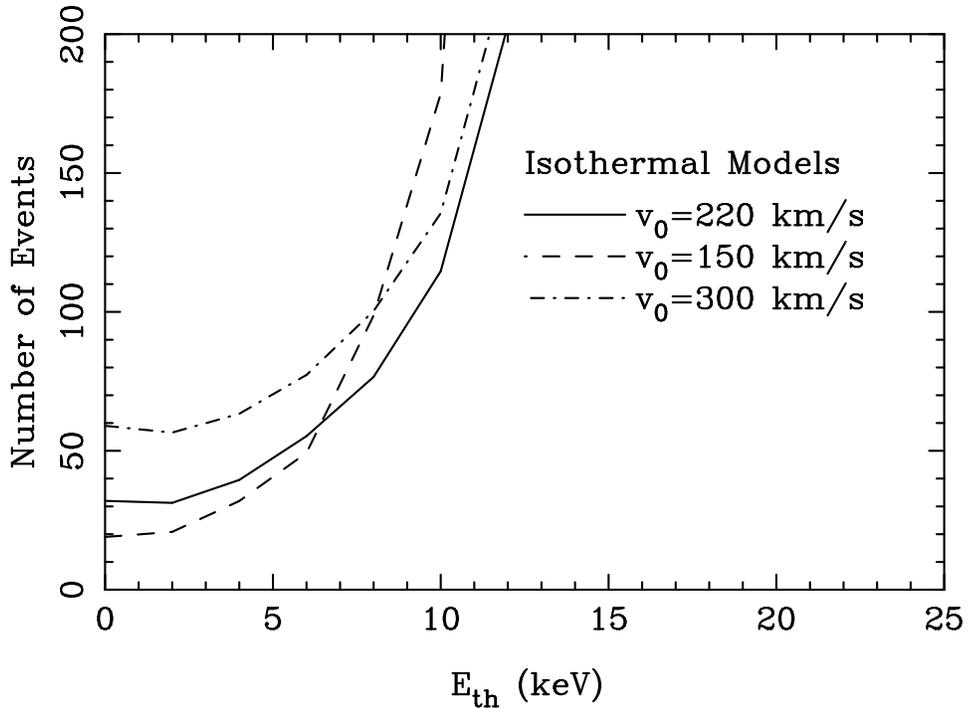}\hss}
  \caption{The number of signal events required as a function of threshold
    energy to distinguish an isothermal distribution of WIMPs in the halo
    from a flat background.  Here we have considered three values for the
    dispersion in the halo, $v_0=220\kmps$ (solid line), $v_0=150\kmps$
    (dashed line), and $v_0=300\kmps$ (long dashed-short dashed line).  See
    the text for a discussion.}
  \label{fig:Nevents-iso}
\end{figure}

The effect of changing the width of the isothermal distribution (changing
$v_0$) is shown in figure~\ref{fig:Nevents-iso}.  As expected from
figure~\ref{fig:models-gamma} at low thresholds the narrower distribution
($v_0=150\kmps$) is more easily distinguished from a flat background than
the broader distributions.  At higher thresholds the tail of the WIMP
distribution becomes important since the number of high energy nuclear
recoils falls off more rapidly for narrower distributions.  Note that it is
still the case that the number of events required for detection is an
increasing function of energy.

\begin{figure}[tbp]
  \leavevmode\hbox to\hsize
  {\hss\epsfig{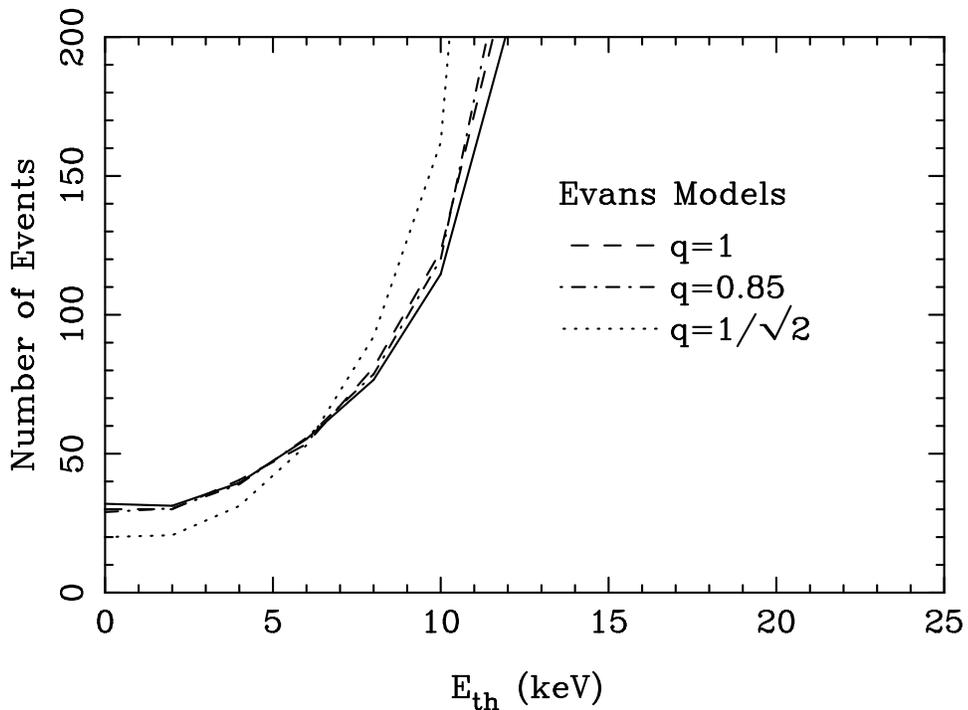}\hss}
  \caption{The number of signal events required as a function of threshold
    energy to distinguish WIMPs in the halo distributed according to an
    Evans model from a flat background.  Here we have considered three
    values for the flattening of the halo, $q=1$ (dashed line), $q=0.85$
    (long dashed-short dashed line), and $q=1/\sqrt2$ (dotted line).  The
    standard $v_0=220\kmps$ isothermal model (solid line) is shown for
    reference.  See the text for a discussion.}
  \label{fig:Nevents-evans}
\end{figure}

The number of events required for the axisymmetric isothermal models are shown in
figure~\ref{fig:Nevents-evans}.  For reference the $v_0=220\kmps$
isothermal model is also included (solid line).  Both the $q=1$ and
$q=0.85$ models are very similar to the $v_0=220\kmps$ isothermal model.
The $q=1/\sqrt2$ (maximal flattening) model is very similar to the
$v_0=150\kmps$ isothermal model.

\begin{figure}[tbp]
  \leavevmode\hbox to\hsize
  {\hss\epsfig{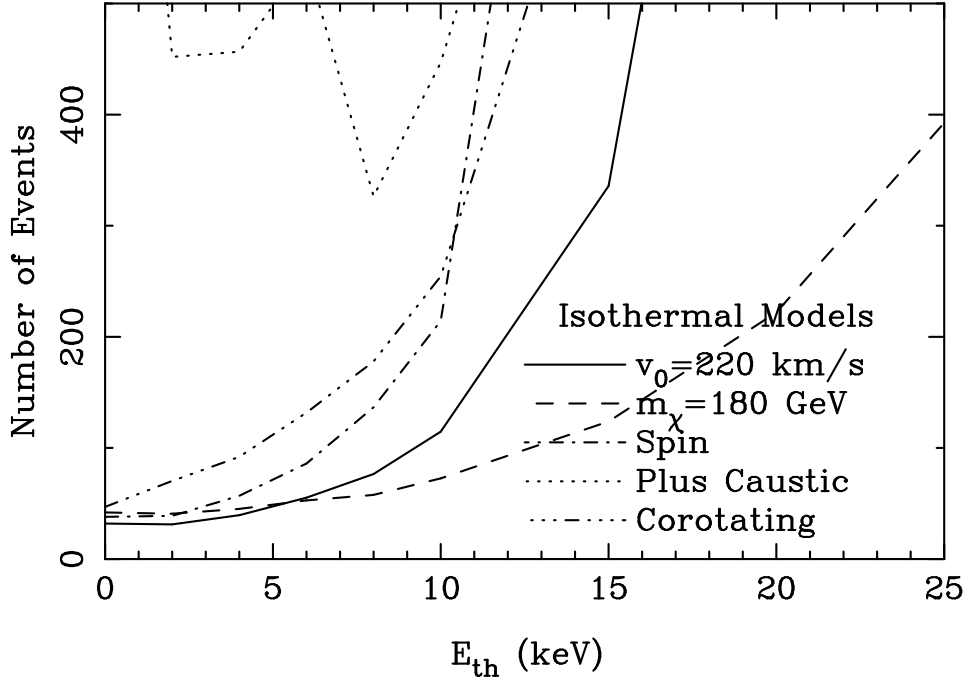}\hss}
  \caption{The number of signal events required as a function of threshold
    energy to distinguish various isothermal distributions of WIMPs in the
    halo from a flat background.  Here we have considered a WIMP of mass
    $m_\chi=180\GeV$ (dashed line), the spin dependent (axial vector)
    interaction (long dashed-short dashed line), the caustic model (with an
    isothermal component, dotted line), and a corotating model
    (dashed-dotted line). The standard $v_0=220\kmps$
    isothermal model (solid line) is shown for reference.  See the text for
    a discussion.}
  \label{fig:Nevents-iso-other}
\end{figure}

In order to explore the general effects of spin-dependence, WIMP mass, and
rotation, we display in figure~\ref{fig:Nevents-iso-other} the predicted
event distributions for a $v_0=220\kmps$ isothermal model, while allowing
these other parameters to vary.  From the figure we see that the spin
interaction using the form factor presented earlier leads to a steeper rise
in the number of events. This is due to the steeper fall off of the
spin-dependent form factor with energy compared to the spin independent
form factor, and will thus be nucleus-dependent.  A larger mass WIMP leads
to more energy transfered to the nucleus, thus the $m_\chi=180\GeV$ WIMP
requires fewer events at high threshold than the lighter $m_\chi=60\GeV$
WIMP does.  A (net) corotating halo is more difficult to distinguish than
the isothermal sphere as expected.  However, it is not as difficult as
might be expected since we have cut out some, but not all, of the WIMPs
from the forward direction.  This shows up more strongly as the threshold
is increased.

It is important to note that all of the above distributions require roughly the same
number of events in order to distinguish a signal from a flat background.  If this
were generic, then low-threshold detectors sensitive to roughly 30-50 events would be
guaranteed to unambiguously identify a WIMP signal, even if a distinction between
halo models might not be easy.   Unfortunately, however, two of the models we have
considered here, both of which exist in the literature as candidates for our
Galactic halo, will require far more events in order to disentangle the signal
from a flat background.  For the reasons mentioned earlier, it is reasonable to
expect that  the distributions that arise from N-body simulations may share these
features.  

First, we consider the caustic distribution.  
As is expected from figure~\ref{fig:angular-caustic}, this most closely resembles
a flat distribution and thus requires the most number of events.  Also not
surprisingly it is a strong function of threshold energy.  As the threshold is
increased some of the velocity peaks do not contribute detectable nuclear recoils
thus greatly changing the shape of the recoil distribution.  We note that the number
of events required for detection is not a monotonic function.  Thus the detection of
WIMPs in a caustic halo will depend sensitively on the mass of the WIMP and  the
threshold of the detector.  A range of thresholds would need to be probed.  In
practice also having energy information about the event will most likely prove to be
extremely important to differentiate WIMP events from background events in this case.

\begin{figure}[tbp]
\vbox{
  \leavevmode\hbox to\hsize
  {\hss\epsfig{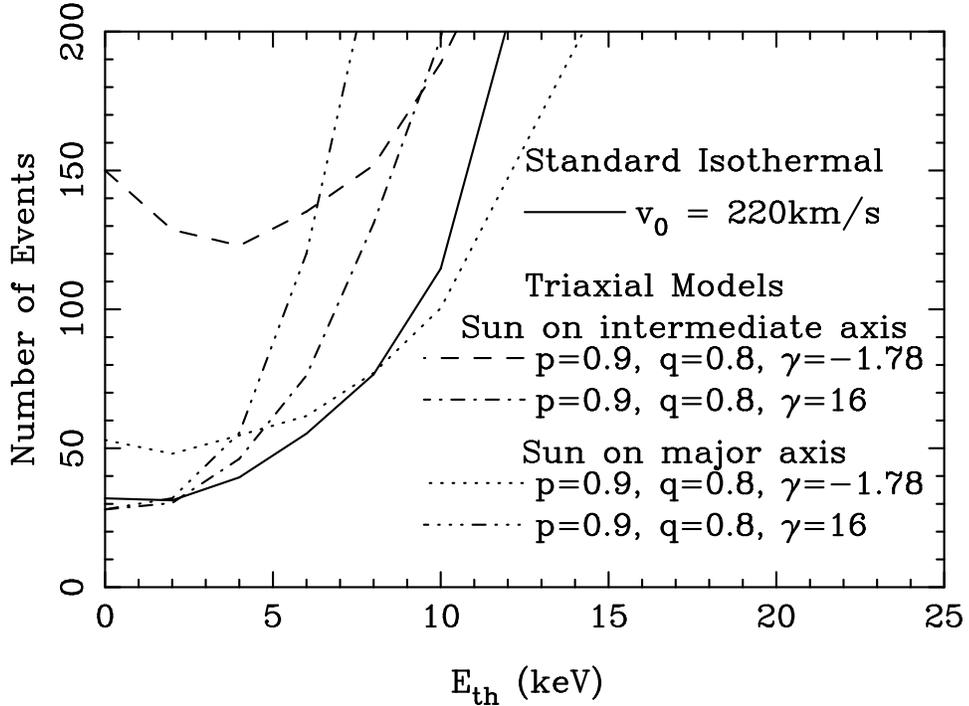}\hss}
  \caption{The number of signal events required as a function of threshold
    energy to distinguish a triaxial distribution of WIMPs in the halo from
    a flat background.  Here we have considered a the Sun on the
    intermediate axis with $\gamma=-1.78$ (dashed line ) and $\gamma=16$
    (long dashed-short dashed line) and the Sun on the major axis with
    $\gamma=-1.78$ (dotted line) and $\gamma=16$ (dashed-dotted line).  The
    standard $v_0=220\kmps$ isothermal model (solid line) is shown for
    reference.  See the text for a discussion.}
  \label{fig:Nevents-tri}
}
\end{figure}

Next we consider the triaxial logarithmic ellipsoidal model.  As expected
from figure~\ref{fig:angular-tri} most of the models are not too different
from the standard isothermal sphere~\ref{fig:Nevents-tri}.  However, when
the Sun is on the intermediate axis and we have a radial anisotropy
($\gamma=-1.78$) approximately 120--150 events are required.  This model
shows the importance of exploring a broad class of halos and parameters in
order to assess the requirements of a a detector with angular resolution.

In the above results we have assumed fairly accurate angular resolution
(bins of about 9 degrees in width were used).  However a real detector may
not be able to achieve this level of angular resolution.  To explore this
we consider the case of only having knowledge of the forward to backward
asymmetry.  In this case the events fall into one of two bins.  In our
likelihood analysis we compare two binomial distributions; one for the
data with fraction of events in the forward bin as our probability and the
other for the flat background where the probability is a half.

\begin{figure}[tbp]
  \leavevmode\hbox to\hsize
  {\hss\epsfig{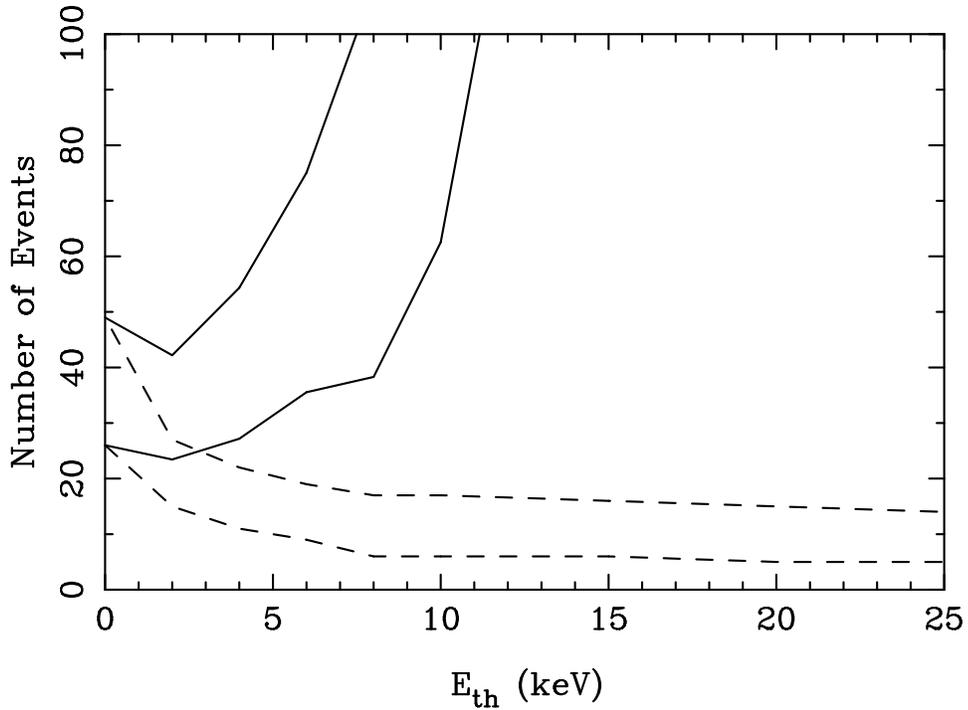}\hss}
  \caption{The number of signal events required as a function of threshold
    energy to distinguish an isothermal distribution of WIMPs in the halo
    with $v_0=220\kmps$ from a flat background using only the forward to
    backward ratio.  See figure~\protect\ref{fig:Nevents-iso220} and the
    text for more details.}
  \label{fig:Nevents-RfRb-iso220}
\end{figure}

In figure~\ref{fig:Nevents-RfRb-iso220} we show the number of signal events 
needed to differentiate a WIMP signal from a flat background at the 95\%
confidence level 95\% of the time when only two angular bins are employed
(forward and backward).  This figure is the counterpart of
figure~\ref{fig:Nevents-iso220}.  Even in this case of relatively little
angular information few events are needed.  For $S/N=1$ if $\Eth<5\keV$
then $N_{\rm events} \approx 40\hbox{--}65$.  It is interesting to note
that thresholds slightly above zero give a somewhat more discernible
signal in this case.

\begin{figure}[tbp]
  \leavevmode\hbox to\hsize
  {\hss\epsfig{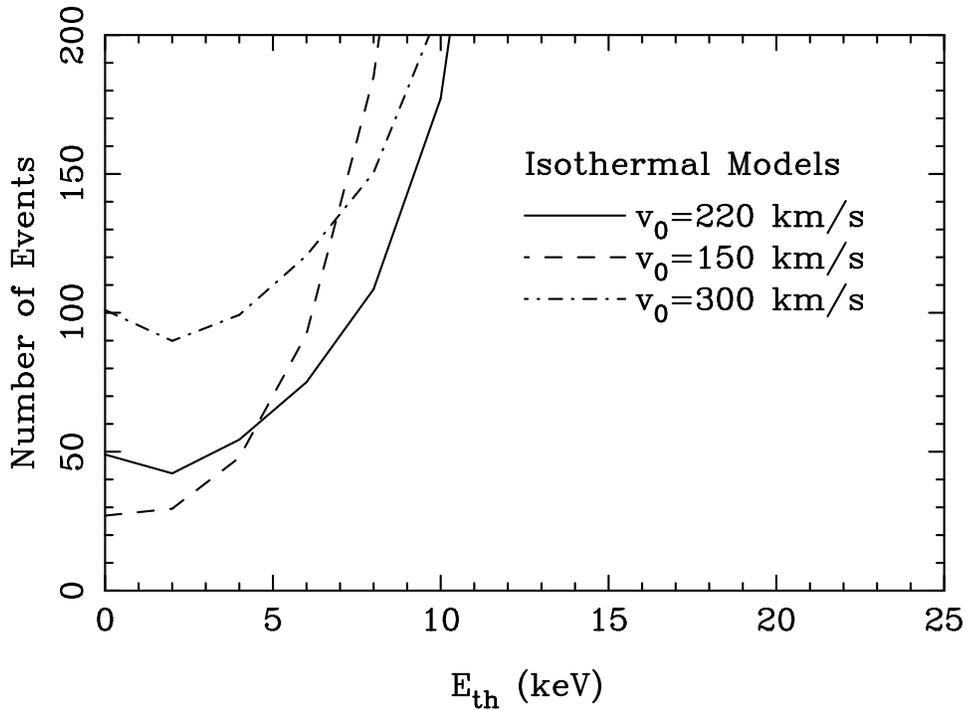}\hss}
  \caption{The number of signal events required as a function of threshold
    energy to distinguish an isothermal distribution of WIMPs in the halo
    from a flat background using only the forward to backward ratio.  Here
    we have considered three values for the dispersion in the halo,
    $v_0=220\kmps$ (solid line), $v_0=150\kmps$ (dashed line), and
    $v_0=300\kmps$ (long dashed-short dashed line).  See
    figure~\protect\ref{fig:Nevents-iso} and the text for more
    details.}
  \label{fig:Nevents-RfRb-iso}
\end{figure}

\begin{figure}[tbp]
  \leavevmode\hbox to\hsize
  {\hss\epsfig{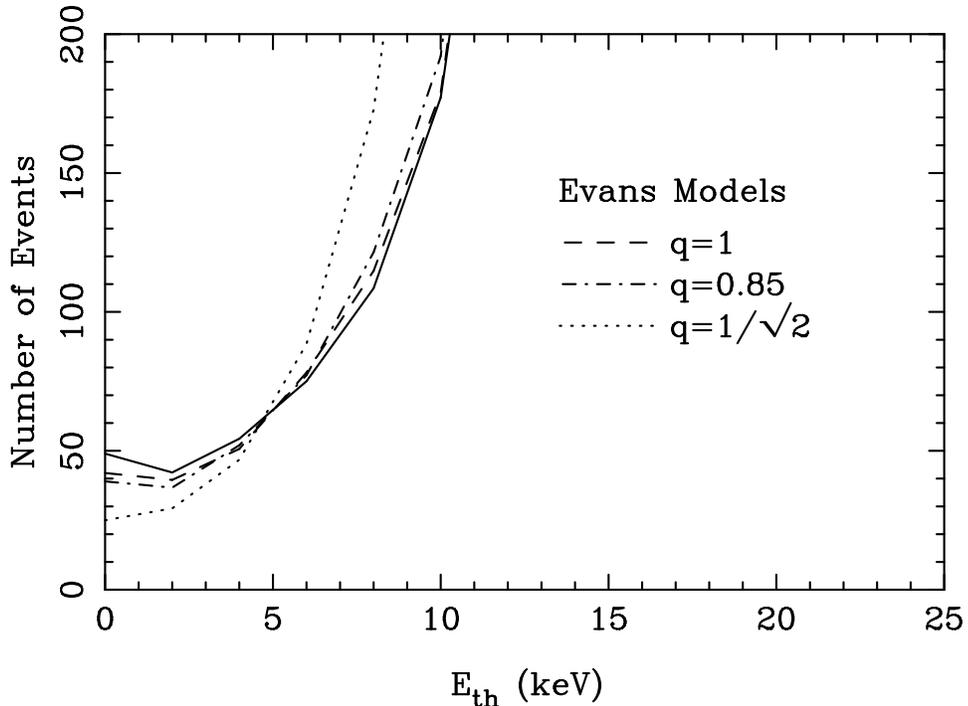}\hss}
  \caption{The number of signal events required as a function of threshold
    energy to distinguish WIMPs in the halo distributed according to an
    Evans model from a flat background using only the forward to backward
    ratio.  Here we have considered three values for the flattening of the
    halo, $q=1$ (dashed line), $q=0.85$ (long dashed-short dashed line),
    and $q=1/\sqrt2$ (dotted line).  The standard $v_0=220\kmps$ isothermal
    model (solid line) is shown for reference.  See
    figure~\protect\ref{fig:Nevents-evans} and the text for more details.}
  \label{fig:Nevents-RfRb-evans}
\end{figure}

\begin{figure}[tbp]
  \leavevmode\hbox to\hsize
  {\hss\epsfig{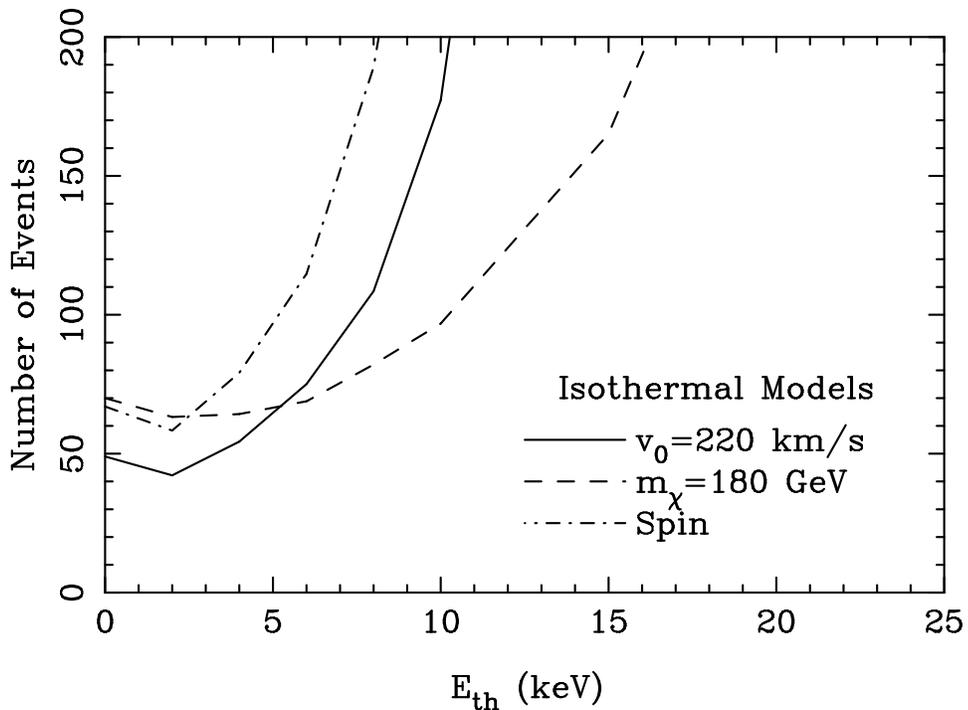}\hss}
  \caption{The number of signal events required as a function of threshold
    energy to distinguish various isothermal distributions of WIMPs in the
    halo from a flat background using only the forward to backward ratio.
    Here we have considered three values for the flattening of the halo,
    $q=1$ (dashed line), $q=0.85$ (long dashed-short dashed line), and
    $q=1/\sqrt2$ (dotted line).  The standard $v_0=220\kmps$ isothermal
    model (solid line) is shown for reference.  See
    figure~\protect\ref{fig:Nevents-iso-other} and the text for more
    details.}
  \label{fig:Nevents-RfRb-iso-other}
\end{figure}

\begin{figure}[tbp]
  \leavevmode\hbox to\hsize
  {\hss\epsfig{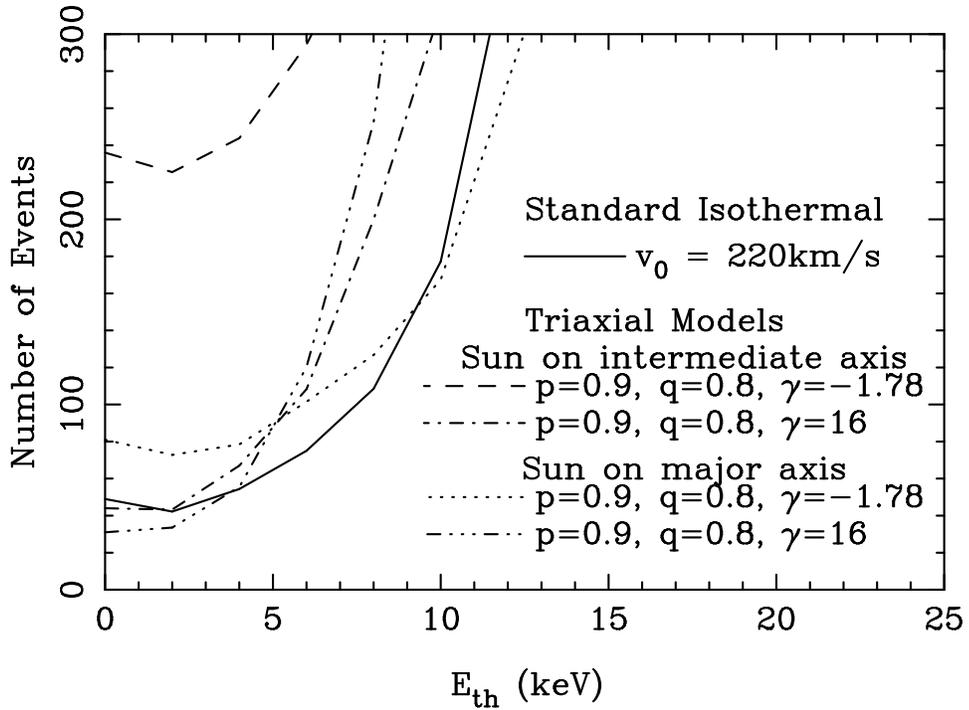}\hss}
  \caption{The number of signal events required as a function of threshold
    energy to distinguish a triaxial distribution of WIMPs in the halo from
    a flat background using only the forward to backward ratio.  Here we
    have considered a the Sun on the intermediate axis with $\gamma=-1.78$
    (dashed line ) and $\gamma=16$ (long dashed-short dashed line) and the
    Sun on the major axis with $\gamma=-1.78$ (dotted line) and $\gamma=16$
    (dashed-dotted line).  The standard $v_0=220\kmps$ isothermal model
    (solid line) is shown for reference.  See
    figure~\protect\ref{fig:Nevents-tri} and the text for a discussion.}
  \label{fig:Nevents-RfRb-tri}
\end{figure}

The results for the other models considered in this work are shown in
figures~\ref{fig:Nevents-RfRb-iso}--\ref{fig:Nevents-RfRb-tri}.  These
figures should be compared to
figures~\ref{fig:Nevents-iso}--\ref{fig:Nevents-tri}.  In most cases we see
the slight dip (or at least very slow rise) in the number of events needed
around $\Eth\approx 2\keV$.  We note that the result for the caustic model
is not shown.  As previously mentioned the caustic model most closely
resembles a flat distribution.  This shows up most prominently here where
we have limited the angular information available.  In this case the number
of events needed with $S/N=1$ at $\Eth=0\keV$ is about 5200.  Again most of
the triaxial models are not that different than the isothermal sphere.
However the radially anisotropic model with the Sun on the intermediate
axis again requires the most number of events.  Here approximately 250--300
events are required for $\Eth<5\keV$.

The DRIFT detector in its current form will only obtain information about
the recoil in two directions.  One will be along the axis of the Sun's
motion through the Galaxy ($z$-axis in our coordinates) and the other
either the $x$ or $y$-axis.  Having information along the $z$-axis is
necessary to have any ability to identify WIMPs.  Since the recoil
distribution is essentially independent of $\phi$ (or equivalently the $x$, 
$y$-plane, see figure~\ref{fig:Nevents-iso220}) having information on these two
axes is almost the same as having the full angular distribution.  We have
calculated the number of events required to identify a WIMP signature in
this case and the results are identical to those discussed above for the
isothermal and axisymmetric models.  Naturally in the case of the caustic model
more events are required since this model has structure in the $x$,
$y$-plane.

This simple study of the DRIFT's capabilities assumed that we could
identify the recoil tracks regardless of their orientation.  In practice if 
we do not have full angular resolution then some tracks will not be
resolved.  For example, if we are projecting out the $x$-axis any event
that has a large component along the $x$-axis and a small component along
the $y$-axis will appear as a short track and will thus be difficult to
identify.  To accurately model DRIFT it would be necessary to take this
track identification into account along with other detector characteristics 
(threshold, etc.) Furthermore it may not be possible to measure the
direction of the track, only its axis.  In a detector such as DRIFT that
relies on ionization to measure the recoil direction, it is difficult to
tell where a track begins and ends.  To make this distinction we must rely
on the ionization (energy deposition) as a function of the energy of the
recoiling nucleus.  The extent to which this can be used to determine
directionality will be crucial for determining the ability of a detector to
identify a WIMP signal.  We will consider these details in future work.  

\section{Conclusions}

Our results are useful when considering the construction of the next
generation of WIMP detectors sensitive to the angular distribution of WIMP 
scattering
events.  As we have described, this sensitivity will provide, in principle,
a key extra signature needed to unambiguously demonstrate that an observed
signal is indeed WIMP-related, and one which will require both far fewer events
and also be subject to fewer systematic uncertainties than exist in a search for
simple annual modulations.

The formalism we have developed allows one to calculate the angular distribution of
recoil events, $dR/d\Omega$, and the distribution in both energy and angle,
$d^2R/dQd\Omega$.  This formalism accommodates arbitrary velocity space WIMP
distributions and correctly employs the full motion of the Earth around the Sun and
the Sun through the Galaxy.  As such, it can be used as an input to any detector
Monte Carlo, allowing a determination of the predicted angular distribution of
events for any halo model.   

While we hope to utilize our analysis to examine various specific 
detectors possibilities in the future, we have thus far derived various
generic predictions for the range of events that will be required in an idealized
detector, given our present uncertainty in the actual Galactic halo WIMP
distribution.  We feel this is an important first step in order to set
the scale of detector parameters that may be required to adequate exploit
angular sensitivity. 

In particular, we have calculated the recoil distribution from
WIMP scattering and the expected number of events necessary to distinguish
a WIMP signal from a background with a 95\% confidence level for
various analytic halo models; isothermal models with $v_0 = 150\kmps$,
$220\kmps$, and
$300\kmps$; Evans models with flattening $q=1$, $0.85$, and $1/\sqrt2$, and
rotating halo models.  In 
all such cases the number of events required is quite low, typically
$50\hbox{--}70$ for $\Eth<5\keV$ when the full angular distribution is
measured.  Even if only the forward to backward ratio is measured
relatively few events are required, typically $70\hbox{--}110$ for
$\Eth<5\keV$ for these models.  This study considered WIMPs with masses
$m_\chi=60\GeV$ and $180\GeV$ and included both spin dependent and
independent interactions.   We note that spin-dependent interactions may
require lower thresholds in order to have sufficient number of events to 
distinguish a WIMP signal.

The isothermal and Evans models are both axisymmetric.  As extreme
asymmetric models we also studied the case of caustics (peaks) in the
velocity distribution, and a set of triaxial halo models.  The former
distribution should be considered a first pass at considering more
realistic halo distributions arising from hierarchical accretion in N-body
simulations.  In this case the resulting recoil spectrum is very similar to
a flat background and thus is rather difficult to distinguish. Here the
number of events required ranges from $300\hbox{--}600$ for
$\Eth=2\hbox{--}10\keV$ and depends rather sensitively on the threshold.
For the triaxial model most sets of parameters lead to limits similar to
the isothermal sphere.  Only the case of the Sun on the intermediate axis
with a radial anisotropy ($\gamma=-1.78$) are the requirements raised to
120--150 events for $\Eth<5\keV$.

We have also performed a preliminary study of a DRIFT-like detector.  Here we
projected out one of the directions perpendicular to the direction of
motion of the Sun since DRIFT will only have resolution in two directions.
Here we found that the results can be as robust as having full
angular resolution for axisymmetric halo models.

In future work we plan to refine our analysis in several ways.  First, we will
exploit the output of existing N-body simulations to explore the consequences for
the expected angular signals.   Next, we plan to carry out  an analysis similar to
the present analysis for specific proposed detectors.  This will allow
us to explore the very important issue of utilizing energy as well as angular
sensitivity, but it will also require us to adequately model the expected energy
dependence of various detector backgrounds.   Finally, after completing these
analyses, we hope to explore exactly how many events may be needed to distinguish
between halo models, in order to determine how an eventual WIMP detection may shed
light on the unknown features of our Galactic halo, and thus on important issues
associated with galaxy formation and evolution.  

If a WIMP signal is ultimately observed in terrestrial detectors
 it will provide one of the most important observations that has ever been
made in particle
physics and cosmology.   
It is thus worthwhile considering in
advance not only how one might best unambiguously disentangle such a signal from
backgrounds, but also how one might then use this signal as a probe of astrophysics.
Our results provide a useful first step in this direction.

\vskip 0.1in
We thank P. Sikivie, J. Martoff, and B. Moore for useful discussions
about their research results, and J. Heo for his contributions to our
initial investigations.

\begin{table}
\vbox{
\caption{Velocity flows of dark matter from the Caustic model (for
$h=0.75$, $\epsilon =0.28$, $j_{\rm max}=0.25$ Sikivie
model~(Sikivie, private communications).
Velocities are given in the rest frame of the Galaxy. }
\label{tab:caustic}
\begin{tabular}{ccccc}
Flow    &
  $\rho$\tablenotemark[1]
   & $v_x$\tablenotemark[2]
  & $v_y$\tablenotemark[2] & $v_z$\tablenotemark[2] \\
\tableline 1 & $0.4$ & $140$ & $0$ & $\pm600$ \\
2 & $0.9$ & $250$ & $0$ & $\pm500$ \\
3 & $2.0$ & $350$ & $0$ & $\pm395$ \\
4 & $6.1$ & $440$ & $0$ & $\pm240$ \\
5 & $9.6$ & $440$ & $\pm190$ & $0$ \\
6 & $3.0$ & $355$ & $\pm290$ & $0$ \\
7 & $1.9$ & $295$ & $\pm330$ & $0$ \\
8 & $1.4$ & $250$ & $\pm350$ & $0$ \\
9 & $1.0$ & $215$ & $\pm355$ & $0$ \\
10 & $1.1$ & $190$ & $\pm355$ & $0$ \\
11 & $0.9$ & $170$ & $\pm355$ & $0$ \\
12 & $0.8$ & $150$ & $\pm350$ & $0$ \\
13 & $0.7$ & $135$ & $\pm345$ & $0$ \\
14 & $0.6$ & $120$ & $\pm340$ & $0$ \\
\end{tabular}
\tablenotetext[1]{In units of  $10^{-26}\;{\rm g/cm^3}$.}
\tablenotetext[2]{In units of $\rm km/s$.}
}
\end{table}

\begin{thebibliography}{99}

\bibitem{review}
G. Jungman, M. Kamionkowski, and K. Greist, Phys. Rep. {\bf 267}, 195 (1996).

\bibitem{cdms}
R.Abusaidi, {\it etal}. (CDMS collaboration), Phys. Rev. Lett. {\bf 84},
5699 (2000).

\bibitem{idm2000}
Proceedings, Third International Conference on the Identification
of Dark Matter, York, England, ed. N. Spooner et al., World Scientific,
to appear.

\bibitem{dama}
R.Bernabei, {\it etal}., Phys. Lett. B {\bf 480}, 23 (2000).

\bibitem{annual-variation}
A.K. Drukier, K. Freese, and D.N. Spergel, Phys. Rev. {\bf D33} 3495 (1986).

\bibitem{spergel}
D.N. Spergel, Phys. Rev. {\bf D37}, 1353 (1988).

\bibitem{angulardet}
C.J. Martoff, {\it etal}., Phys. Rev. Lett. {\bf 76}, 4882 (1996).

\bibitem{tpc}
D.P. Snowden-Ifft, C.J. Martoff, and J.M. Burwell, Phys. Rev. Lett.,
submitted, astro-ph/9904064 (1999);
M.J. Lehner, {\it etal}., astro-ph/9905074 (1999).

\bibitem{copietal}
C.J. Copi, J. Heo, and L.M. Krauss, Phys. Lett. B {\bf 461}, 43 (1999).

\bibitem{smith} 
P.F. Smith and J.D. Lewin, Phys. Rep. {\bf 187} 203 (1990).

\bibitem{evans}
N.W. Evans, Mon. Not. R. Astron. Soc. {\bf 260}, 191 (1993).

\bibitem{kamionkowski}
M. Kamionkowski and A. Kinkhabwala, Phys. Rev. {\bf D57}, 3256 (1998).

\bibitem{triaxial}
N.W. Evans, C.M. Carollo, and P.T. de Zeeuw, astro-ph/0008156 (2000).

\bibitem{caustic}
P. Sikivie, I.I. Tkachev, and Y. Wang, Phys. Rev. {\bf D56}, 1863 (1997).

\bibitem{Gespin}
M. Ted Ressel, {\it etal}., Phys. Rev. {\bf D48}, 5519 (1993).

\bibitem{lynden-bell}
D. Lynden-Bell, MNRAS {\bf 136}, 101 (1967).

\end{thebibliography}
\end{document}